\documentclass[prd,superscriptaddress,showpacs,twocolumn,nofootinbib,floatfix,amsmath,amssymb]{revtex4-1}
\usepackage{mathtools}
\usepackage[usenames,dvipsnames]{xcolor}
\usepackage{color}
\usepackage{braket}
\usepackage{bm}
\usepackage{graphicx}
\usepackage{hyperref}

\renewcommand{\vec}[1]{\bm{\mathrm{#1}}}
\def\mat#1{\bm{\mathrm{#1}}}
\def\op#1{\hat{#1}}
\def\opvec#1{\op{\vec{#1}}}

\newcommand{\ii}{\mathrm{i}}

\begin{document}

\title{Thermal amplification of field-correlation harvesting}
\author{Eric G. Brown}
\email{e9brown@uwaterloo.ca}
\affiliation{Department of Physics and Astronomy, University of Waterloo, Waterloo, Ontario N2L 3G1, Canada}

\begin{abstract}
We study the harvesting of quantum and classical correlations from a hot scalar field in a periodic cavity by a pair of spatially separated oscillator-detectors. Specifically, we utilize non-perturbative and exact (non-numerical) techniques to solve for the evolution of the detectors-field system and then we examine how the entanglement, Gaussian quantum discord, and mutual information obtained by the detectors change with the temperature of the field. While (as expected) the harvested entanglement rapidly decays to zero as temperature is increased, we find remarFkably that both the mutual information $and$ the discord can actually be \emph{increased} by multiple orders of magnitude via increasing the temperature. We go on to explain this phenomenon by a variety of means, and are able to make accurate predictions of the behavior of thermal amplification. By doing this we also introduce a new perspective on harvesting in general and illustrate that the system can be represented as two dynamically decoupled systems, each with only a single detector. The thermal amplification of discord harvesting represents an exciting prospect for discord-based quantum computation, including its use in entanglement activation.
\end{abstract}

\maketitle

\section{Introduction}

Entanglement, a form of quantum correlation without classical analogue, has long been understood to be a key resource in many of the procedures developed in quantum computation and information \cite{Horodecki09}. While the mutual information generically quantifies the total amount of correlation between subsystems that can potentially be useful for computational tasks, entanglement has proven to have the purely quantum nature that gives an advantage over classical computation.

In recent years there has been an explosion of interest in the so called quantum discord \cite{Ollivier01, Henderson01}, as well as similar measures \cite{Modi12}, that purport to quantify quantum correlations that entanglement generally overlooks. That is, a separable (non-entangled) state may still possess quantum correlations in the sense that its joint-measurement statistics cannot be described by classical probability theory. In addition to being of theoretical interest  \cite{Modi12,Girolami13,Streltsov13,Brown13-2,Madhok13}, discord has also received considerable attention regarding its potential as a quantum computational resource \cite{Knill98,Dakic10,Chuan12,Madhok13-2,Almedia13,Gu13}. While the full utility of discord is still far from certain, and many of the protocols discovered have been criticised as being either highly construed and not very useful or in reality utilizing entanglement in an unobvious way, there are inarguably several examples of discord being a genuinely useful quantum resource. A prime example of this is the realization in \cite{Piani11,Streltsov11}, and the recent experimental demonstration in \cite{Adesso13}, that quantum discord quantifies the amount of distillable entanglement generated between a system and a measurement apparatus upon performing a local measurement. Here, we will consider a measure called the Gaussian discord \cite{Adesso10, Giorda10}, which is specifically suited to Gaussian continuous-variable systems; more information about this specific measure will be provided in later sections.

In another sector of literature there has been an increasing interest in what some now dub entanglement harvesting, or extraction, from a quantum field \cite{Reznik05,Braun05,Leon09,Steeg09}. This scenario involves two spatially separated model detectors made to locally interact with a quantum field (typically a massless scalar field in the vacuum state) and the result is that these detectors may become entangled through this process. This is even true for causally disconnected detectors (their interaction time is shorter than the light-travel time between them) and this is explained by the fact that a vacuum field is already spatially entangled \cite{Summers87,Srednicki93}, and the detectors are simply harvesting some of the entanglement already present. While an interesting theoretical discovery, there has been little hope that this phenomenon may be used as a practical tool for generating entanglement over large distances because the amount of entanglement obtained is typically minuscule. That said, there has very recently been a proposal for a reusable and sustainable entanglement ``farming" procedure that may be able to overcome this limitation \cite{Edu2013}.

In addition to entanglement, we study in this paper both the mutual information and the quantum discord locally harvested from a field. Excepting \cite{Borrelli12}, such studies have not previously been undertaken in the literature. Here, we furthermore generalize the state of our field to consider both the vacuum as well as thermal states of varying temperature, and obtain rather exciting results (see below). To perform this study we utilize a continuous variable approach introduced in \cite{Brown13}, and similar to the tools used in \cite{Dragan11,Bruschi13}, in which the detectors placed in the field are modeled as harmonic oscillators. By doing this we are able to easily solve for the evolution of the detectors-field system both non-perturbatively and, in the case at hand, exactly (non-numerically). This represents a considerable advantage over the qubit-based detector model most widely used in the literature, in which perturbation theory is required. To utilize the oscillator-detector approach we will consider a massless scalar field that is enclosed in a periodic cavity. While limiting ourselves to a cavity field may feel restrictive, the fact is that an experimental demonstration or computational utilization of this phenomenon is likely to take place in a cavity setting \cite{Sabin11,Sabin12}, and thus understanding how it works in a cavity is of importance.

To be clear, the ability to locally extract discord does \emph{not} follow from the same physics that allows the well-known creation of discord through local channels \cite{Strel}. Such creation generally comes at the cost of reduced mutual information, and discord \emph{cannot} be locally created from a product state. In the correlation harvesting scenario we consider two detectors that are initially uncorrelated, and both their discord and mutual information are increased from zero due to their local interactions with the field.

Studying the effects of thermal fluctuations on our ability to harvest correlations is clearly an important part of understanding how to actually perform such a procedure in a laboratory setting, since at least some level of noise will always be present in realistic setups. Generally, it is understood that noise experienced by one's system has a detrimental effect on the correlations present in that system due to the decoherence that such noise typically induces. One should similarly expect that thermality in a field will reduce the ability of detectors to harvest correlations from the field; the increased thermal fluctuations will cause the detectors to become more mixed and furthermore it is known that the spatial entanglement present in a field reduces with temperature \cite{Anders08,Ferraro08}. However, there have also been several studies demonstrating that discord is typically more robust than entanglement against the detrimental effects of noise \cite{Werlang09,Wang2010,Datta2009,Brown2012,Doukas2013}, and we might therefore hope that in a harvesting scenario discord will hold out better.

Our findings are in fact much better than this hope would warrant. While we find that, as expected, the harvested entanglement decays rapidly to zero as field temperature is increased, we simultaneously find that both the mutual information \emph{and} the Gaussian quantum discord between the detectors actually \emph{increase} with field temperature. This increase is in fact quite drastic, with an improvement of multiple orders of magnitude being achievable. This increased extraction of both classical and quantum correlations is what we will refer to as ``thermal amplification".

After presenting our results, we go on to discuss multiple ways in which this surprising result can be explained and understood. Of particular importance is a detailed analysis that takes advantage of the translational invariance of the periodically-identified field. We illustrate that our system can be decomposed into two dynamically decoupled systems, each with only a single detector, and for which the evolution behavior is easily understood. By this we are able to explain both the thermal amplification of mutual information and discord, as well as the decay of entanglement, in terms of the correlating properties of passive Gaussian transformations. Using this, we find that we can also make accurate predictions for the strength of thermal amplification as a function of the system parameters. The new perspective on correlation-harvesting that we introduce in this explanation is something that we feel is interesting and worthy of study in its own right.

The thermal amplification of discord is especially surprising, as discord is purported to measure purely quantum correlations. This finding introduces the exciting possibility of using a cavity field to non-locally generate what is an appreciable amount of discord, which can then be used in discord-based quantum computing. This is especially so since, apparently, experimentalists need not be concerned with keeping their cavity very cold. For example, as was mentioned above, the discord and related measures quantify the amount of distillable entanglement that can be activated by local interaction with an ancilla system \cite{Piani11,Streltsov11,Adesso13}. Thus it seems that while thermal fluctuations are indeed detrimental to entanglement directly, they may indirectly be of great benefit to its generation. We let ourselves ponder the possibility that this type of thermal amplification will lead to the development of what may be called ``noise-assisted quantum computation".

After the initial preprint of this manuscript we became aware that similar results to ours have been observed in \cite{Paz}. This paper, however, considers a different scenario (in their case the oscillator detectors are \emph{directly} coupled), uses very different techniques, and contains very different interpretations from what we present here. Furthermore, in addition to presenting some interesting and counterintuitive results, it is our goal here to explain the phenomenon of thermal amplification from several independent points of view. In this way, we hope to aid in the possibility of utilizing this effect in experiment and in practical application.

Our paper is organized as follows. In Sect. \ref{model} we describe the system of detectors and a cavity field that we will consider, and we overview the continuous-variable techniques introduced in \cite{Brown13} that are used to solve for the evolution of this system. We also describe how to compute the three correlation measures that we will be analyzing (the logarithmic negativity as a measure of entanglement, the mutual information, and the Gaussian quantum discord) between the two detectors. In Sect. \ref{sectResults} we will present our primary results and in Sect. \ref{explain} we will discuss and explain the results via several perspectives. We finish with concluding remarks in Sect. \ref{conclusions}. We also include an appendix, \ref{app}, that goes into detail on using the material introduced in Sect. \ref{explain} to make accurate predictions on the parameter-dependence of thermal amplification.

Throughout this paper we will maintain unitless quantities, with $c=\hbar=k_B=1$, and with all entropic quantities measured in bits (i.e. $\log \equiv \log_2$).

\section{The model, evolution, and correlation}  \label{model}

Here we describe the scenario under consideration, overview the continuous variable technique that we will use to solve for the system evolution, as well as define the correlation measures that we consider. For a more complete explanation of these techniques the reader is referred to \cite{Brown13}. We will be heavily utilizing the formalism of Gaussian quantum mechanics; the unfamiliar reader can refer to one of many resources available in the literature (for example \cite{Adesso07}) explaining this formalism. Unless otherwise cited, everything presented here is fully explained in one of these two papers.

Our scenario consists of a massless scalar field in a one-dimensional periodic waveguide of length $L$ (we will also refer to this as the cavity). The field modes are solutions to the wave equation with periodic boundary conditions and normalized with respect to the Klein-Gordon inner product \cite{BirrelDavies84}. In practice we will need to also apply a UV cutoff to the field so that the number of field modes is finite. We have been sure to include enough field modes such that adding additional modes does not noticeably change the results.

We will let the field include $N$ right-moving modes and $N$ left-moving modes. We will find it most convenient to work in the Schr\"odinger picture, in which case our cavity field takes the simple form
\begin{align} \label{field}
	\hat{\phi}(x)=\sum_{n=-N}^{N} \frac{1}{\sqrt{4 \pi |n|}} \left( e^{\ii k_n x} \op{a}_n + e^{-\ii k_n x} \op{a}_n^\dagger \right),
\end{align}
where the wave-numbers are $k_n= 2 \pi n /L$. Here $\hat{a}_n^\dagger$ and $\hat{a}_n$ are the creation and annihilation operators of the field modes. It should also be noted that, as is commonplace in the literature, we exclude the zero-mode in Eq. (\ref{field}); subsequent sums over field modes will be understood to also have this exclusion.

We consider what happens when we let a pair of harmonic oscillators (i.e. the detectors) locally interact with this field. We let the detectors each have characteristic frequencies of $\Omega$, and we  place them in the cavity at positions $x_1$ and $x_2$ where they are left to remain stationary. We let the detectors start in their ground states, although we could just as easy consider other states. For the initial state of the field, we shall consider both the vacuum state and more generally thermal states of various temperatures.

By modeling the detectors as oscillators we are putting them on mathematically equivalent grounds to the field modes. Specifically, both detectors and modes are described as continuous-variable bosonic degrees of freedom. We label the quadrature operators of these modes by $(\op{q}_i,\op{p}_i)$, where $i$ runs over all detectors and field modes, and they satisfy the standard canonical commutation relations: $[\op q_i,\op p_j]=\ii \, \delta_{ij}$. We package these operators in a phase-space vector of the form
\begin{align}  \label{xvec}
	\opvec{x}=(\hat{q}_{d1},\hat{p}_{d1}, \hat{q}_{d2}, \hat{p}_{d2},\hat{q}_{-N},\hat{p}_{-N}, \dots, \hat{q}_N,\hat{p}_N)^T,
\end{align}
where the first $4$ entries correspond to the detectors and the remaining $4N$ entries to the field modes. These operators are related to their corresponding annihilation and creation operators by
\begin{align}
	\hat{q}_i=\frac{1}{\sqrt{2}}(\hat{a}_i+\hat{a}^\dagger_i), \;\;\;\; \hat{p}_i=\frac{\ii}{\sqrt{2}}(\hat{a}_i^\dagger-\hat{a}_i).
\end{align}
This relation holds for both the detectors and the field modes.

Given such a continuous-variable ensemble, we will consider the state of our detector-field system to be a Gaussian states \cite{Adesso07}. Specifically, as explained in \cite{Brown13}, we only need to consider zero-mean Gaussian states. Such a state is fully described by the covariance matrix $\mat{\sigma}$, the entries of which are given by

\begin{align} \label{covmat}
	\sigma_{i j}=\langle \hat{x}_i \hat{x}_j+\hat{x}_j \hat{x}_i \rangle.
\end{align}

As we will see, describing continuous-variable physics in a finite dimensional phase-space rather than an infinite dimensional Hilbert space is what allow us to solve the evolution of the system non-perturbatively. In order to preserve the Gaussianity of the system, however, the evolution must be generated by a no more than quadratic Hamiltonian \cite{Schumaker86}. We must therefore choose such a detector-field interaction Hamiltonian. Fortunately, many of the interactions of interest in physics are of this form. Here, we will choose to use the monopole-monopole type coupling that has been so widely utilized in Unruh-DeWitt detector models \cite{DeWitt80}. The corresponding interaction Hamiltonian (again, in the Schr\"odinger picture) between the field and two stationary detectors at positions $x_1$ and $x_2$ is
\begin{align} \label{Hint}
	\hat{H}_\text{int}= \lambda_1 (\hat{a}_{d1}+\hat{a}_{d1}^\dagger)\hat{\phi}(x_1)+ \lambda_2 (\hat{a}_{d2}+\hat{a}_{d2}^\dagger)\hat{\phi}(x_2),
\end{align}
where $\hat{a}_{d1}$ and $\hat{a}_{d2}$ are the annihilation operators of the detectors and the numbers $\lambda_1$ and $\lambda_2$ are the coupling strengths, which will here be considered to be equal, $\lambda_1=\lambda_2=\lambda$. Note that the case in which detectors are not stationary is easily represented; one simply replaces $x_{1,2}$ with the detector positions as a function of time \cite{Brown13}.

Working in the Schr\"odinger picture means we must consider the total Hamiltonian when generating evolution, including both the free and interacting parts: $\hat{H}=\hat{H}_\text{free}+\hat{H}_\text{int}$. The free Hamiltonian contains parts coming  from both the detectors and from the field:
\begin{align}  \label{Hfree}
            \hat{H}_\text{free}=\Omega \hat{a}_{d1}^\dagger \hat{a}_{d1}+\Omega \hat{a}^\dagger_{d2} \hat{a}_{d2}+&\sum_{n=-N}^{N} \omega_n \hat{a}_n^\dagger \hat{a}_n \\
	=\frac{\Omega}{2}(\op{p}_{d1}^2+\op{p}_{d2}^2+\op{q}_{d1}^2+\op{q}_{d2}^2)+&\sum_{n=-N}^N \frac{\omega_n}{2}(\op{p}_{n}^2+\op{q}_{n}^2),  \nonumber
\end{align}
where $\Omega$ is the frequency of the detectors and $\omega_n=|k_n|$ since the field is massless. Note that we have ignored any constant additions to the Hamiltonian, since these will not have any impact on the evolution. 

The unitary evolution generated by a quadratic Hamiltonian can be represented in phase-space by a symplectic transformation. In terms of the covariance matrix, which fully describes the state of our system, such a transformation as a function of time takes the form
\begin{align}  \label{covmatevol}
	\mat{\sigma}(t)=\mat{S}(t)\mat{\sigma}_0 \mat{S}(t)^T,
\end{align}
where $\mat{\sigma}_0$ represents the initial state and $\mat{S}(t)$ is a symplectic matrix. Such a matrix is defined by the condition
\begin{align}
	\mat{S} \mat{\Omega} \mat{S}^T=\mat{S}^T \mat{\Omega} \mat{S}=\mat{\Omega},
\end{align}
where $\mat{\Omega}$ is the symplectic form, defined as
\begin{align} \label{symform}
	\mat{\Omega}=\bigoplus_i
	\begin{pmatrix}
		0 & 1 \\
		-1 & 0
	\end{pmatrix},
\end{align}
where $i$ runs through all degrees of freedom (both detectors and field modes).

The goal is now to find what symplectic evolution $\mat{S}(t)$ is generated by a given quadratic Hamiltonian $\op H$. The procedure for doing this is described in \cite{Brown13}, and here we reiterate the result. Any (in general time-dependent) quadratic Hamiltonian can be represented as a Hermitian matrix $\mat{F}(t)$ on phase-space such that
\begin{align}
	\hat{H}(t)=\opvec{x}^T \mat{F}(t) \opvec{x}.
\end{align}
From this, the symplectic evolution $\mat{S}(t)$ generated by $\hat{H}(t)$ solves a Schr\"odinger-type equations of the form
\begin{align} \label{EOM}
	\frac{d}{dt} \mat{S}(t)=\mat{\Omega} \mat{F}^\text{sym}(t) \mat{S}(t),
\end{align}
where $\mat{F}^\text{sym}=\mat{F}+\mat{F}^T$ and $\mat{\Omega}$ is the $2N+2$ mode symplectic form. 

When the two detectors are stationary inside the cavity, as we consider here, then the total Hamiltonian is time independent. In this case the solution to Eq. (\ref{EOM}) is trivially given by
\begin{align} \label{soln}
	\mat{S}(t)=\exp (\mat{\Omega} \mat{F}^\text{sym} t).
\end{align}
More generally, in time-dependent scenarios, Eq. (\ref{EOM}) must be solved numerically.

Working in the Schr\"odinger picture we must include both the free and interaction Hamiltonians in the evolution equation. The symmetrized Hamiltonian matrix can thus be decomposed into the form $\mat{F}^\text{sym}=\mat{F}^\text{sym}_\text{free}+\mat{F}^\text{sym}_\text{int}$. For our particular scenario, as given by Eqs. (\ref{Hint},\ref{Hfree}), we have explicitly $\mat{F}^\text{sym}_\text{free}=\text{diag}(\Omega,\Omega,\Omega,\Omega, \omega_N, \omega_N, \omega_{N-1},\omega_{N-1}, \dots, \omega_N,\omega_N)$ and
\begin{align}
	\mat{F}^\text{sym}_\text{int}=2\lambda
	\begin{pmatrix}
		\mat{0}_{4} & \mat{X} \\
		\mat{X}^T & \mat{0}_{4N}
	\end{pmatrix},
\end{align}
where $\mat{0}_n$ is the $n \times n$ matrix of zeros, and
\begin{widetext}
\begin{eqnarray}
	\mat{X} \equiv
	\begin{pmatrix}
		\frac{\cos( k_{-N}x_1)}{\sqrt{4 \pi N}} & \frac{-\sin( k_{-N}x_1)}{\sqrt{4\pi N}} & \frac{\cos (k_{-N+1} x_1)}{\sqrt{4 \pi(N-1)}} & \frac{-\sin (k_{-N+1}x_1)}{\sqrt{4 \pi (N-1)}} & \dots & \frac{\cos (k_N x_1)}{\sqrt{4\pi N}} & \frac{-\sin (k_N x_1)}{\sqrt{4 \pi N}} \\
		0 & 0 & 0 & 0 & \dots & 0 & 0 \\
		\frac{\cos( k_{-N}x_2)}{\sqrt{4 \pi N}} & \frac{-\sin( k_{-N}x_2)}{\sqrt{4\pi N}} & \frac{\cos (k_{-N+1} x_2)}{\sqrt{4 \pi(N-1)}} & \frac{-\sin (k_{-N+1}x_2)}{\sqrt{4 \pi (N-1)}} & \dots & \frac{\cos (k_N x_2)}{\sqrt{4\pi N}} & \frac{-\sin (k_N x_2)}{\sqrt{4 \pi N}} \\
		0 & 0 & 0 & 0 & \dots & 0 & 0
	\end{pmatrix}.
\end{eqnarray}
\end{widetext}

Upon obtaining the symplectic evolution matrix, the state of the system as a function of time from initial state $\mat{\sigma}_0$ is described by $\mat{\sigma}(t)=\mat{S}(t) \mat{\sigma}_0 \mat{S}(t)^T$. It is important to stress that the computation of $\mat{S}(t)$ in no way depends on the initial state. This is therefore very convenient when applied to studies of different initial states, such as in this paper, because the dependence on the initial state can be studied without recomputing the evolution.

Here we are going to consider the detectors and the field to be initially uncorrelated, although this is not necessary for the formalism. Further, we will initialize the detectors in their ground states and initialize the field in a thermal Gibbs state of temperature $T$. This includes the $T=0$ limit, which is the vacuum state. Note that by vacuum and thermal we are referring to such states as defined with respect to the free Hamiltonian, Eq. (\ref{Hfree}). The covariance matrix of the ground/vacuum state of system of oscillators/modes is simply given by the identity matrix. Specifically, for the pair of detectors and the field:
\begin{align}
	\mat{\sigma}^{(d)}_\text{ground}=\mat{I}_4, \;\;\;\;\;\;\; \mat{\sigma}^{(f)}_\text{vac}=\mat{I}_{4N}.
\end{align}
More generally, a thermal state of the field is given by the covariance matrix
\begin{align} \label{thermalstate}
	\mat{\sigma}^{(f)}_\text{therm}= \bigoplus_{n=-N}^N
	\begin{pmatrix}
		\nu_n & 0 \\
		0 & \nu_n
	\end{pmatrix},
\end{align}
where   
\begin{align}   \label{thermeigs}
	\nu_n=\frac{\exp{\omega_n \beta}+1}{\exp{\omega_n \beta}-1}, \;\;\;\; \beta \equiv 1/T,
\end{align}
are the symplectic eigenvalues of a thermal state (see \cite{Adesso07}, for example, for an explanation of symplectic eigenvalues). This means that the initial state we use is of the form
\begin{align}
	\mat{\sigma}_0=\mat{\sigma}^{(d)}_\text{ground} \oplus \mat{\sigma}^{(f)}_\text{therm}.
\end{align}


After evolution, the state of our system will take the generic form
\begin{align}
	\mat{\sigma}=
	\begin{pmatrix}
		\mat{\sigma}^{(d)} & \mat{\gamma} \\
		\mat{\gamma}^T & \mat{\sigma}^{(f)}
	\end{pmatrix},
\end{align}
where $\mat{\sigma}^{(d)}$ and $\mat{\sigma}^{(f)}$ are the $4\times 4$ and $4N \times 4N$ covariance matrices describing the reduced states of the detectors and the field, respectively. The matrix $\mat \gamma$ stores the information regarding correlations between the detectors and the field. By examining the state of the detectors, $\mat{\sigma}^{(d)}$ we can compute correlation measures between the detectors such as the logarithmic negativity \cite{Plenio05}, the mutual information, and the Gaussian quantum discord \cite{Adesso10,Giorda10}. The reader is referred to these papers for a full description of these measures in Gaussian states, as we will only very briefly introduce them here.

To begin, the von Neumann entropy of a general Gaussian state with symplectic eigenvalues $\{\nu_i\}$ is given by
\begin{align} \label{entropy}
	S(\mat{\sigma})=\sum_{i=1}^N f(\nu_i),
\end{align} 
where
\begin{align}   \label{ffunction}
	f(x)=\frac{x+1}{2} \log \left( \frac{x+1}{2}\right)-\frac{x-1}{2}\log \left(\frac{x-1}{2}\right).
\end{align}

Since we wish to study the correlations extracted by the two detectors, we are interested in the correlations contained in a two-mode Gaussian state. The covariance matrix of the detector-detector state that we obtain upon evolution will be of the generic form
\begin{align} \label{twomode}
	\mat{\sigma}^{(d)}=
	\begin{pmatrix}
		\mat{\sigma}_1 & \mat{\gamma}_{1 2} \\
		\mat{\gamma}_{1 2}^T & \mat{\sigma}_2
	\end{pmatrix},
\end{align}
where the $2\times 2$ covariance matrices $\mat {\sigma}_1$ and $\mat{\sigma}_2$ describe the reduced states of detectors-$1$ and $2$. The matrix $\mat{\gamma}_{12}$ contains information about the correlations between the two detectors. For example, the detectors will be uncorrelated (i.e. in a product state) iff all entries of $\mat{\gamma}_{12}$ are zero.

The mutual information between the detectors, which quantifies the total correlation, then follows the usual form:
\begin{align}  \label{mutinf}
	I=S(\mat{\sigma}_1)+S(\mat{\sigma}_2)-S(\mat{\sigma}^{(d)}).
\end{align}
As just stated, $I=0$ iff  $\mat{\gamma}_{12}=\mat{0}$.

If $\mat{\sigma}^{(d)}$ is a pure state (meaning that its symplectic eigenvalues are $\nu_1=\nu_2=1$) then the entanglement between detectors is fully characterized by the reduced entropy $S(\mat{\sigma}_1)=S(\mat{\sigma}_2)$. This will not be the case in our scenario however, and we will therefore use instead the logarithmic negativity as our entanglement measure \cite{Plenio05}. Let us define the quantities $\alpha=\det \mat{\sigma}_1$, $\beta=\det \mat{\sigma}_2$,  $\gamma= \det \mat{\gamma}_{12}$, and $\delta=\det \mat{\sigma}^{(d)}$. The logarithmic negativity between the detectors is then given by
\begin{align} \label{logneg}
	E_N=\max (0, -\log \tilde{\nu}_-),
\end{align}
where $\tilde{\nu}_-$ is the smaller of the state's partially transposed symplectic eigenvalues (these are \emph{not} generally the same as the ordinary values). This can be computed from
\begin{align}
	2\tilde{\nu}_-^2=\tilde{\Delta}-\sqrt{\tilde{\Delta}^2-4 \delta},
\end{align}
where $\tilde{\Delta}=\alpha+\beta-2 \gamma$. In the case of two-mode Gaussian states the logarithmic negativity vanishes iff $\mat{\sigma}^{(d)}$ is a separable state.

Lastly, we will consider the Gaussian quantum discord $D$ of a two-mode Gaussian state \cite{Adesso10, Giorda10}. Quantum discord is a measure of the purely quantum part of correlations obtained by subtracting off the classical correlations from the mutual information  \cite{Ollivier01, Henderson01}. The discord between two subsystems is zero iff the correlation structure can be described by a classical probability distribution. When computing discord, one is required to perform an optimization over local measurements. This generally makes such a computation very difficult for high-dimensional systems. When working with Gaussian states, the best that has been achieved in this regard is to optimize over the restricted set of Gaussian measurements (namely, measurements that preserve Gaussianity). This gives the quantity known as Gaussian discord, and it permits an analytic solution in the case of two-mode Gaussian states that is given by \cite{Adesso10}
\begin{align} \label{discord}
	D(1:2)=f(\sqrt{\beta})-f(\nu_1)-f(\nu_2)+f(\sqrt{E}),
\end{align}
where $\nu_1$ and $\nu_2$ are the symplectic eigenvalues of $\mat{\sigma}^{(d)}$ and
\begin{widetext}
\begin{eqnarray}
E&=& \left\{  
\begin{array}{l}
\frac{{2 \gamma^2+\left(-1+\beta \right) \left(-\alpha+ \delta \right)+2 |\gamma| \sqrt{\gamma^2+\left(-1+\beta \right) \left(-\alpha+ \delta \right)}}}{{\left(-1+\beta \right){}^2}} ~~~\text{for}\quad\left (\delta-\alpha \beta  \right) {}^2 \le \left (1 +   \beta \right) \gamma^2 \left (\alpha + \delta \right),\\ \\
\frac{{\alpha \beta-\gamma^2+\delta-\sqrt{\gamma^4+\left(-\alpha \beta+\delta \right){}^2-2 \gamma^2 \left(\alpha \beta+ \delta \right)}}}{{2 \beta}}\quad \;\;\;\;\;\;\;\;\;\, \hbox{otherwise}.
\end{array} \right.
   \nonumber
\end{eqnarray}
\end{widetext}
This is the case in which the optimized measurement has been considered on system-2. In general, the discord is not symmetric with respect to this choice: $D(1:2) \neq D(2:1)$. However, in our our scenario we will find that the two detectors are symmetric under exchange, and thus here we will have $D\equiv D(1:2)=D(2:1)$.

It should be noted that although there is circumstantial evidence that Gaussian measurements are actually optimal for Gaussian states \cite{Giorda12}, there is as of yet no proof of this, and so it is possible that the Gaussian discord may slightly overestimate the true value of discord in general.

\section{Results}   \label{sectResults}

We can now present the primary results of this paper, which were obtained using the formalism above. We consider what occurs when two detectors, initially in their ground states, are injected into a cavity field that is either in its vaccum state or in a thermal state. We then track the evolution of correlation measures between the detectors, including logarithmic negativity $E_N$,  Eq. (\ref{logneg}), mutual information $I$, Eq. (\ref{mutinf}), and Gaussian quantum discord $D$, Eq. (\ref{discord}). The results obtained for logarithmic negativity follow exactly as would be intuitively expected, and thus it will not be our primary focus here. Rather it is the mutual information and discord that display unexpected behavior and will take up the majority of this paper. In this section we will merely present our results, and in the following section we will go on to give several explanations for them and discuss.

In short, our primary result is as follows: thermality of the field can be used to \emph{increase} the amount of non-entanglement correlation that is extracted from the field by the detectors. That is, both the mutual information \emph{and} the quantum discord that are extracted increase as a function of field temperature $T$. On the other hand, the harvested entanglement rapidly vanishes as $T$ is increased, agreeing with intuition.

In the following, we use $r=|x_1-x_2|$ to indicate the distance between the detectors. The results of our calculations are independent of the absolute positions $x_1$ and $x_2$, rather only on their difference $r$, because a periodically identified vacuum or thermal field is translationally invariant.

In all data presented here we use the following parameter values: the length of the cavity is $L=100$, the coupling strength for both detectors is $\lambda=0.05$, the detector frequencies are both $\Omega=40 \pi/L \approx 1.26$ (meaning that they are resonant with the $20^{\text{th}}$ field modes, both right and left-moving), and the number of right and left-moving field modes is $N=80$ (this number was chosen such that further increasing $N$ does not perceivably alter the results).

\begin{figure}[t]
	\centering
        \includegraphics[width=0.38\textwidth]{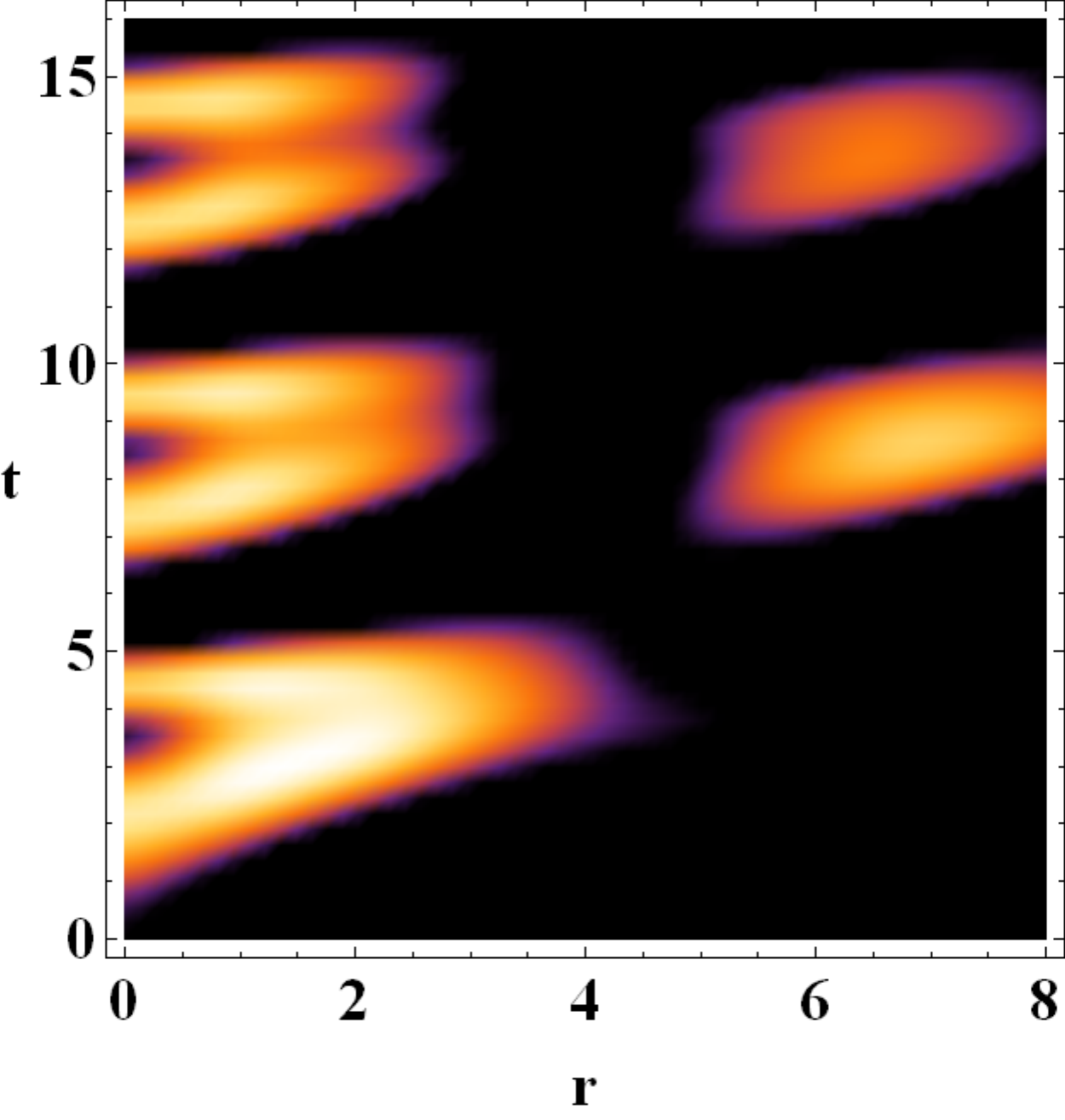}
	\includegraphics[width=0.09 \textwidth]{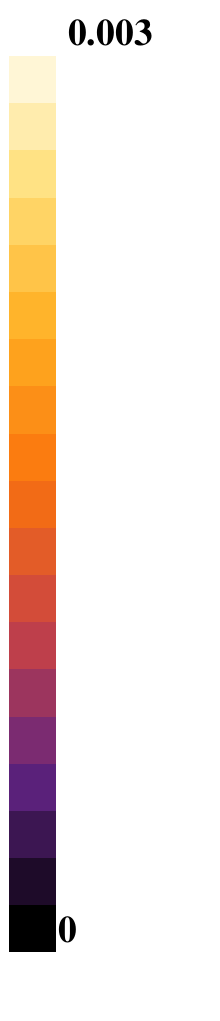}
	\caption{(Color online) The logarithmic negativity between the detectors $E_N$, Eq. (\ref{logneg}), as a function of the distance $r$ between them and the time $t$ of evolution in the case that the field is initiated in its vacuum state.}
        \label{3Dent}
\end{figure}

\begin{figure}[t]
	\centering
    \includegraphics[width=0.45\textwidth]{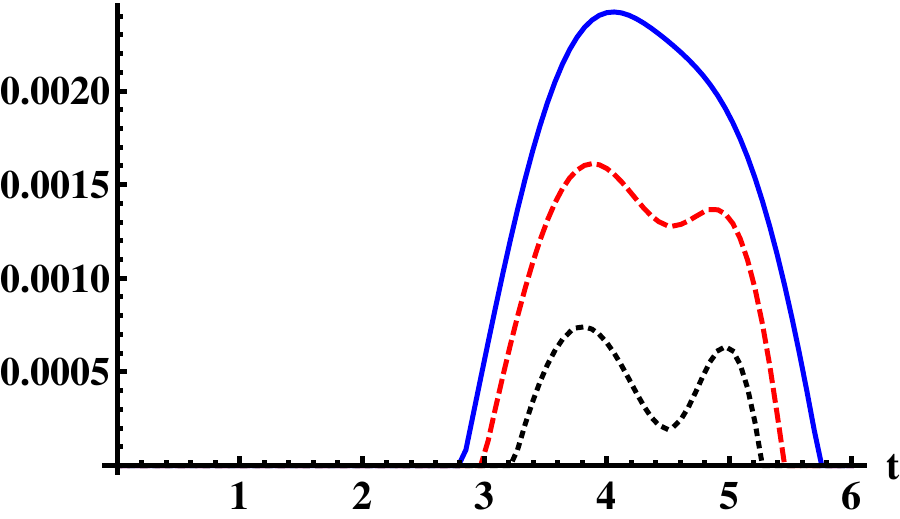}
	\caption{(Color online) The logarithmic negativity between the detectors as a function of time $t$, where $r=3$. We display the data for the cases that the field is initiated in its vacuum state $T=0$ (solid blue line), and in thermal states of temperatures $T=0.1$ (red dashed line) and $T=0.15$ (black dotted line). At a temperature of $T=0.2$ the entanglement remains nonexistent. We observe that the extracted entanglement rapidly decays with field temperature.}
        \label{entdeg}
\end{figure}

Although entanglement will not be our primary focus here, since its behavior follows as expected, for completeness we will include some data regarding its extraction in our scenario. This will also be used as a comparison with the other measures discussed below. We begin by presenting some data for the case that the field starts in the vacuum. In Fig. \ref{3Dent} we plot the logarithmic negativity between the detectors, Eq. (\ref{logneg}), as a function of the distance $r$ between them and the time $t$ of evolution in the case that the field is initiated in its vacuum state. If we compute the same information when the field is instead started in a thermal state, we find that the magnitude of this plot rapidly decays with temperature. This behavior is as expected, and follows one's intuition regarding thermal fluctuations as being a source of decohering noise. As an example, we plot in Fig. (\ref{entdeg}) the logarithmic negativity as a function of time for several different field temperatures. At a distance of $r=3$, we find that any extractable entanglement is completely extinguished by the time the temperature reaches the small value of $T=0.2$.

\begin{figure}[t]
	\centering
        \includegraphics[width=0.38\textwidth]{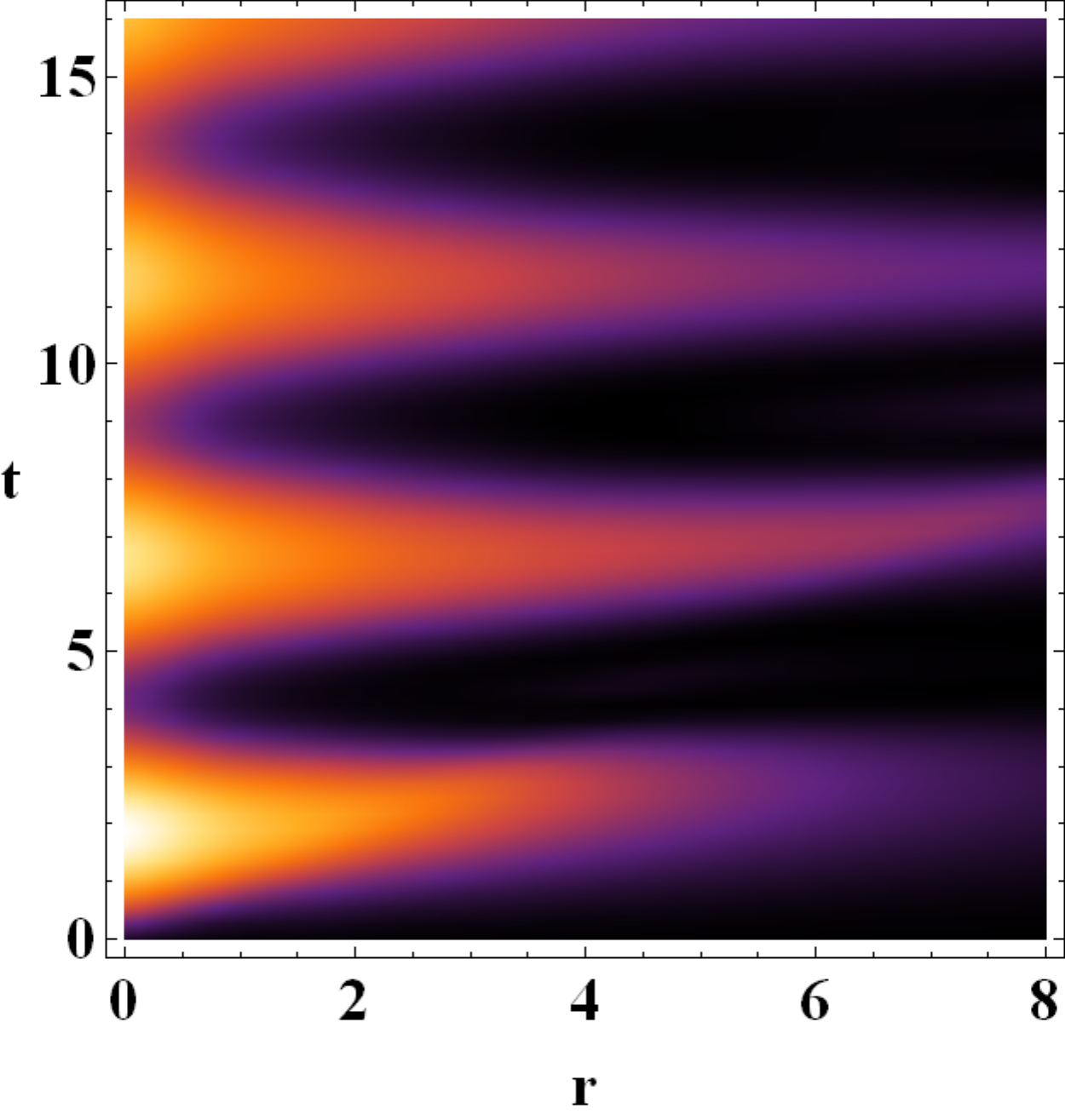}
	\includegraphics[width=0.09 \textwidth]{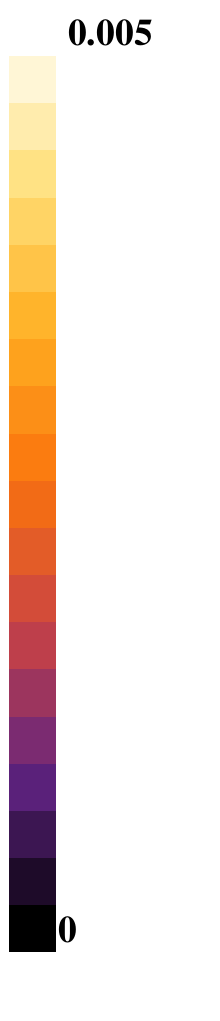}
        \label{3DI}	
	\caption{(Color online) The mutual information between the detectors $I$, Eq. (\ref{mutinf}), as a function of the distance $r$ between them and the time $t$ of evolution in the case that the field is initiated in its vacuum state.}

\end{figure}

\begin{figure*}[t]
	\centering
                 \includegraphics[width=0.32\textwidth]{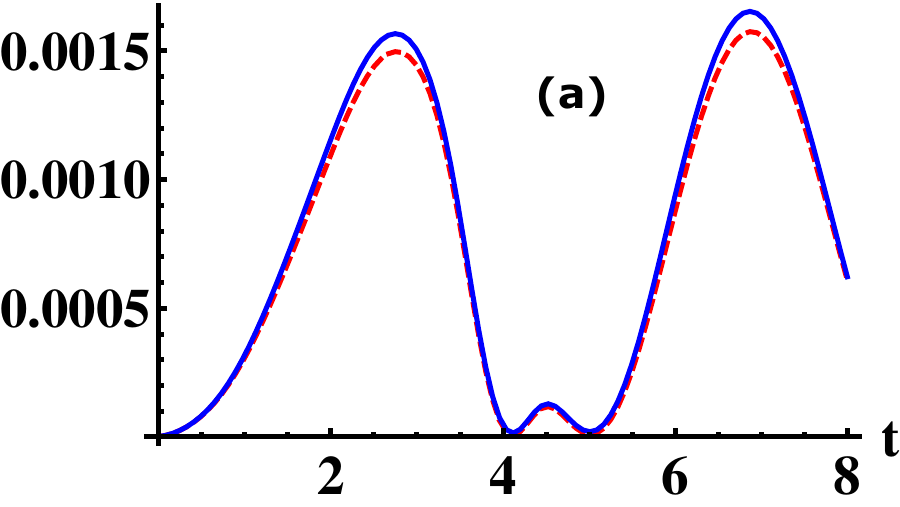}
                 \includegraphics[width=0.32\textwidth]{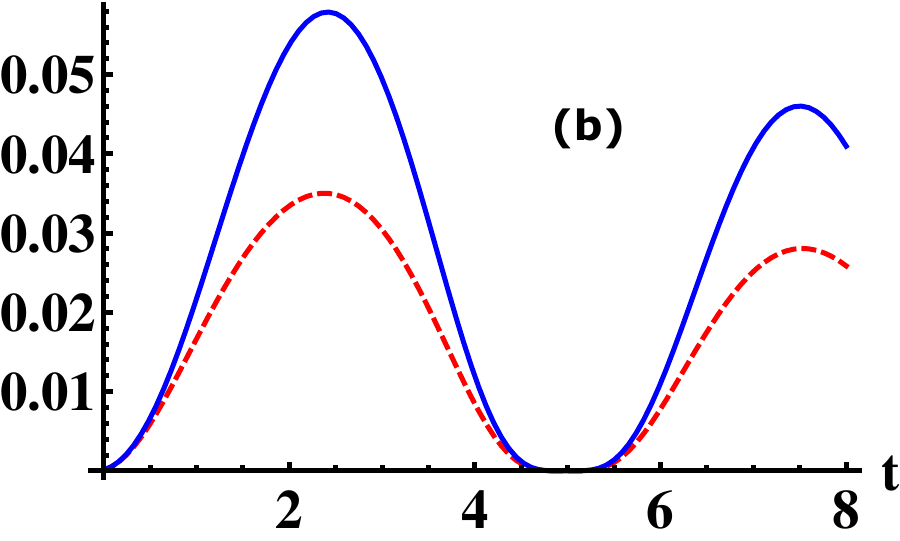}
                  \includegraphics[width=0.32\textwidth]{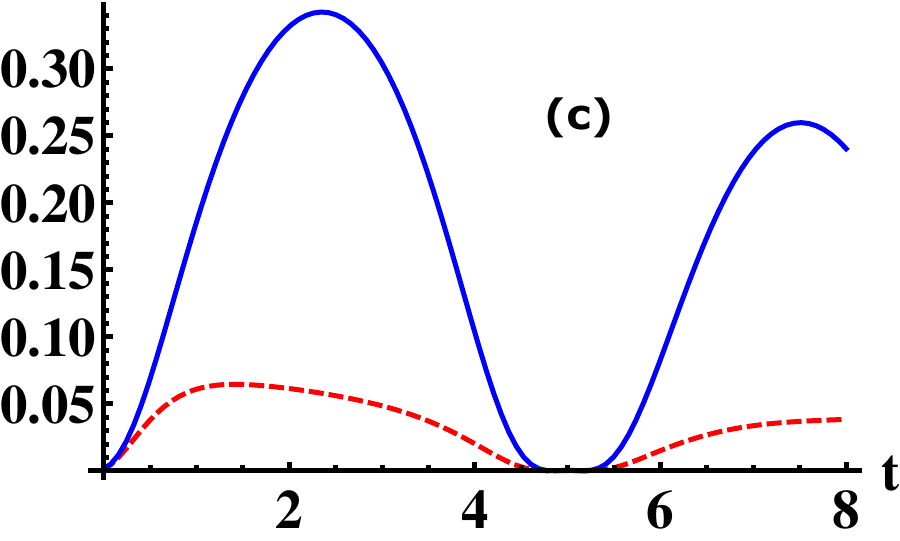}
	\caption{(Color online) The mutual information $I$ (solid blue line) and the Gaussian discord $D$ (dashed red line) between the detectors as a function of time $t$, where $r=4$. The temperature of the field varies in the three plots from (a) $T=0$ (i.e. the vacuum), (b) $T=1$, and (c) $T=10$. We observe that as temperature increases both $I$ \emph{and} $D$ very substantially increase.}
        \label{vst}
\end{figure*}

Moving on, we plot in Fig. \ref{3DI} the mutual information between the detectors, Eq. (\ref{mutinf}), in the same scenario (i.e. with the field initiated in its vacuum state). This clearly displays very different behavior as compared to the logarithmic negativity. We don't bother displaying the Gaussian discord here, Eq. (\ref{discord}), because in this case the discord is only very slightly less than the mutual information (see Fig. \ref{vst}) and thus follows the same behavior. This on its own is an interesting finding: the harvested mutual information consists almost entirely of quantum correlations (at least as quantified by the Gaussian discord). We will see that when considering a thermal field of high temperature this is no longer the case.

Another obvious point to make is that, unlike entanglement, the detectors begin to gain some mutual information (as well as discord) immediately after the interaction is turned on (see Fig. \ref{vst}), and this statement is independent of the distance $r$ between them. Indeed the amount of correlation becomes appreciable far before the light crossing time $t=r$ between the detectors, the time at which they come into causal contact. Such immediate generation of correlations was also seen in \cite{Borrelli12} using a perturbative framework and in the context of the Fermi problem (i.e. when one of the detectors starts in an excited state). It is not overly surprising to find these results. Physically, the field at spatially separated positions is known to be correlated, and so the response of detectors at these positions should be expected to be correlated. While entanglement is clearly a different story, there is no reason to suspect that more general correlations should not begin accumulating immediately after initializing the interaction. Mathematically, it is straightforward to show for small $t$, by expanding Eq. (\ref{soln}) in powers of $t$, that the off-diagonal block $\mat{\gamma}_{12}$ of the detector-detector covariance matrix generically grows as order $t^2$. This implies that the detectors must necessarily have non-zero correlation for any finite time.

We remind the reader that the local harvesting of discord that we observe is \emph{not} the same as the known ability to create discord through local operations \cite{Strel}. Local operations \emph{cannot} increase the mutual information, and often an increase in discord through local operations comes at the cost of an overall reduction in mutual information. Local operations cannot introduce discord into a product state. Here we are clearly seeing a different phenomenon, since both the discord and the mutual information are increasing from zero.

We now go on to present our primary results. Plotted in Fig. \ref{vst} is the mutual information and Gaussian discord between the detectors as a function of time, where the detectors have been placed at a distance $r=4$ away from each other. We display three plots: the first for the case that the field is initialized in its vacuum state (i.e. at temperature $T=0$), the second for the case that the field is initialized in a thermal state at temperature $T=1$, and the third for a temperature of $T=10$. What we observe is that as the field temperature is increased both the obtained mutual information \emph{and} discord are \emph{increased} as well, and by orders of magnitude at that. 
Note that all plots in Fig. \ref{vst} were made using the same symplectic evolution matrix $\mat{S}(t)$ (i.e. it only needed to be solved for once); for each of the three different plots the same $\mat{S}(t)$ was simply applied to a different initial covariance matrix.

\begin{figure}[t]
	\centering
        \includegraphics[width=0.45\textwidth]{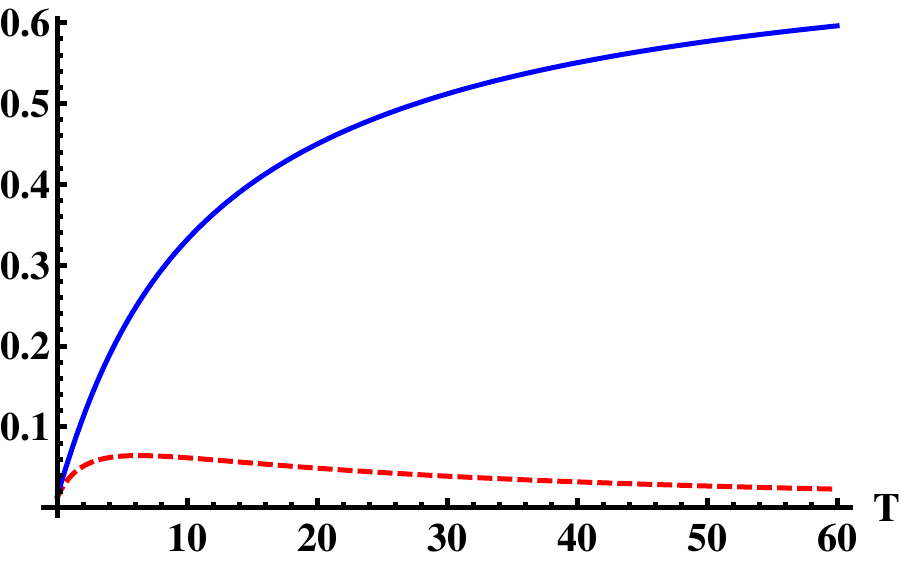}
	\caption{(Color online) The mutual information $I$ (solid blue line) and the Gaussian discord $D$ (dashed red line) between the detectors as a function of field temperature, where $r=4$ and $t=2$.}
        \label{vsT}
\end{figure}

To examine the limits of this behavior, we plot in Fig. \ref{vsT} the mutual information and discord as a function of field temperature $T$ up to very high temperatures, where the detectors were placed at a distance $r=4$ away from each other and left to evolve for a time $t=2$. We can see that, although slowing down, the mutual information continues to increase even at a temperature of $T=60$. The discord, on the other hand, reaches its maximum at approximately $T=6$ before slowly decaying. Even though the amplification of discord (a measure of \emph{quantum} correlations) does eventually cease, this field temperature is relatively very high and well into what one would call a purely classical regime. Nevertheless, it is reassuring to see that thermal decoherence \emph{does} eventually start to hinder the harvested discord.

How do our results change with other parameters of the system, such as the coupling strength and distance? Also, what is the behavior in the long-time limit? In terms of the coupling strength $\lambda$ the harvesting changes as expected: increasing $\lambda$ generically increases the amount of harvested correlations as given by each of our measures. Unfortunately we cannot consider indefinitely large $\lambda$ in our model because the UV modes will become significant in the evolution and one would need to work without a UV cutoff. In the long time limit the mutual information and discord of the detectors do not approach zero, as may have been expected. Rather they continue to oscillate with a characteristic period of $2\pi / \Omega=5$, and with their mean values generically maintained near the values seen in Figs. \ref{vst} and \ref{vsT} for a given temperature. Regarding the dependence on distance $r$, we defer the discussion of this to Sect. \ref{sectCorr} and the Appendix \ref{app}, but the short story is that the mutual information and discord follow the same qualitative behavior with time as well as with temperature (i.e. thermal amplification still occurs) for different distances.

There are several further points to make on these results. We see that as temperature increases the difference between the mutual information and the discord grows greater. In the vacuum case, as we noted above, $I$ and $D$ are nearly equivalent. This implies that the correlations obtained by the detectors are mostly of a quantum nature, with little contribution coming from classical correlations. As the field temperature increases, however, we find that classical correlations begin to take on the dominant role. Nevertheless, for temperatures up to about $T=6$ the harvested discord \emph{does} increase by almost two orders of magnitude from the vacuum case. This is very surprising considering how quickly any extractable entanglement vanishes with increasing temperature. It is known that discord tends to be more robust to decoherence than entanglement \cite{Werlang09,Wang2010,Datta2009,Brown2012,Doukas2013}, but what we have found in this scenario is instead a complete reversal of behavior between the two measures.

This result can be looked at in one of two ways: either one takes this to mean that the Gaussian quantum discord is clearly not an appropriate measure of purely quantum correlations, or we count this as an excellent indication of how quantum computing may still be performed in noisy environments (as the case appears to be, even being enhanced in certain ways). Being optimistic, we will adhere to the latter. Recall that, as discussed in the introduction, it is known that discord can be used to locally activate distillable entanglement between the discordant system and an ancilla  \cite{Piani11,Streltsov11,Adesso13}. Our results therefore indicate that while thermal fluctuations are indeed detrimental to entanglement directly, it may be that they can actually greatly improve its rate of production in an indirect manner.

Before continuing to our discussion, we should briefly relate our scenario and results with that of \cite{Paz}, which we became aware of after our initial preprint. The authors of that paper used a master equation approach and also discovered that the discord between detectors can increase with the temperature of a common environment. However, there is a key difference between their scenario and ours. That is, in \cite{Paz} they considered their detectors to be \emph{directly} coupled to each other in addition to being coupled to a common bath. This is unlike in our scenario, where each detector is only interacting locally with the field. In particular, their study can not be considered an example of harvesting. Interestingly, in their case they discovered that the discord increased monotonically with temperature, asymptoting to a finite value. This is unlike our finding, as in Fig. \ref{vsT}, in which the discord eventually starts to decrease for high enough temperatures. This difference may exactly be due to the additional direct coupling considered in the other paper.

\section{Explanation and discussion}  \label{explain}

In this section we wish to give some further physical and mathematical insight into the surprising results presented above. We do this by first giving a brief discussion of the spatial correlation function of the field and point out that its form is in fact completely consistent with the result of thermal amplification of extracted correlations. We also give some speculation on how our results may also be understood in terms of system-environment entanglement. We then go on to give a large discussion on how this behavior can be understood in terms of the translational invariance of the periodically-identified cavity field. Not only will this allow us to readily predict for what choices of parameters the thermal amplification will be strongest and weakest, it will also reveal an interesting new perspective on the procedure of correlation harvesting in general that we feel is worthy of consideration in its own right. We briefly point out that with this new perspective we can also immediately explain the result of entanglement degradation, as seen in Fig. (\ref{entdeg}), via the known entangling properties of passive Gaussian transformations.

\subsection{Correlation function}   \label{sectCorr}

It is commonly said that thermal states contain no correlations, or at least the amount of correlations present decreases with temperature. This is a rather vague statement, however, and depending on what exactly one means it is demonstrably false. The statement that a thermal state contains $no$ correlations is simply in reference to the different energy modes, which of course are in a product state with respect to each other (as they are also in the vacuum). This does not say anything about spatial correlations however, which are what we are interested in here. While it is certainly true that the entanglement between spatially separated regions decays with temperature \cite{Anders08,Ferraro08}, this does not imply that the same is true for correlations in general.

Our detectors are placed at positions $x_1$ and $x_2$ in the cavity. These detectors become correlated due to the fact that the field fluctuations (both quantum and classical) at these two points are correlated. That is, the detectors are ``measuring" the field at these points. A standard way of determining the degree to which the measurement statistics of two quantum observables $\op{A}$ and $\op{B}$  on separate Hilbert spaces are correlated given some joint state is to compute the correlation function $C(\op{A},\op{B}) \equiv \braket{\op{A}\op{B}}-\braket{\op A}\braket{\op B}$, where the expectation value is taken with respect to whatever state the joint system is in. To clarify, in this definition we are using the simplified notation $\op{A}\op{B}=\op{A} \otimes \op{B}$, $\op{A}=\op{A}\otimes \op{I}$ and $\op{B}=\op{I} \otimes \op{B}$.

If we wish to ask how much the field $\op{\phi}$ is correlated at points $x_1$ and $x_2$, the simplest measure to compute is the correlation function: $C(x_1,x_2)\equiv C(\op{\phi}(x_1),\op{\phi}(x_2))=\braket{\op{\phi}(x_1)\op{\phi}(x_2)}-\braket{\op{\phi}(x_1)}\braket{\op{\phi}(x_2)}$. Of course, in both the vacuum and any thermal state the first moment of the field will vanish (i.e. the Wigner function has zero mean). The correlation function thus takes the form of the equal-time Wightman function \cite{BirrelDavies84}, which due to the translational invariance of the field will only depend on the distance $r=|x_2-x_1|$:
\begin{align}
	C(r)\equiv C(x_1,x_2)=\braket{\op{\phi}(x_1)\op{\phi}(x_2)}.
\end{align}
Note that the condition $C(r)=0$ does \emph{not} necessarily imply the lack of any correlations, however the condition $C(r)>0$ \emph{does} imply the existence of correlations.

We now ask the question of how this quantity changes with field temperature. The answer is that it indeed grows with temperature in qualitative agreement with our results. For example it is known that in free space and in three spatial dimensions the equal-time Wightman function of a massless scalar field in a thermal state of temperature $T$ is \cite{Weldon00}
\begin{align}
	C_{\text{free}}(r)=\frac{T}{4\pi r} \coth (\pi T r).
\end{align}
This correlation function grows monotonically with $T$, and indeed linearly so for large $T$. Of course this is not the correct correlation function in our situation of a cavity field (periodically identified) in one spatial dimension. For us, the correct function is straightforwardly shown to be given by
\begin{align} \label{corrfunc}
	C(r)=\frac{1}{L}\sum_{n>0} \frac{\nu_n}{\omega_n}\cos(\omega_n r),
\end{align}
where the sum is only over positive $n$ and the values $\nu_n$ are the symplectic eigenvalues of the thermal state as given by Eq. (\ref{thermeigs}). The magnitude of this function also grows monotonically with temperature, and thus from this perspective it is not at all surprising that the correlations transfered to the detectors (corresponding to the correlated measurement statistics of the field) should grow with temperature.

The function in Eq. (\ref{corrfunc}) actually gives excellent predictions for how the harvested correlations that we compute directly behave with the distance between detectors. For example we note that for the cavity length of $L=100$ that was used in our results the magnitude of $C(r)$ for high temperatures reaches a minimum (a zero in fact) at a distance of approximately $r \approx 21$. This minimum can be directly seen as a minimum in the extracted correlation when plotted as a function $r$; see the appendix \ref{app}. For larger $r$ beyond this the harvest \emph{increases} until hitting a local maximum at $L/2=50$ (as predicted by Eq. (\ref{corrfunc})). From here any further increase in distance actually corresponds to a decrease in distance, due to the periodicity of the cavity. We note that generally the phenomenon of thermal amplification appears to occur independent of the distance between the detectors, in the sense of a greatly increased discord harvest as compared to the vacuum state value.

\subsection{Relation to system-environment entanglement}

Here we wish to give a more physically insightful interpretation of our results by pointing out the possible connection between thermal amplification and the results presented in the papers \cite{KW, Brown13-2, Madhok13}, regarding the link between discord within a system and the entanglement between that system and its purifying environment. 

In \cite{KW, Brown13-2} it was demonstrated that the discord present in a general bipartite state is deeply related to the entanglement structure in the system's purification. In particular, the discord typically grows with the entanglement between the system and its purifying ancilla, and furthermore the presence of discord requires the presence of both bipartite and genuine tripartite entanglement in the purification. In \cite{Madhok13} the authors considered a coupled, pure $N$-qubit system and then studied a $2$-qubit subsystem from this ensemble. They observed that the discord between these two qubits is completely monotonic with the entanglement entropy between them and the other $N-2$ qubits; this is to be expected from \cite{KW, Brown13-2}. The authors of \cite{Madhok13} also discovered that there is an inverse relation between the $2$-qubit entanglement (using the concurrence) and the $2$-qubit discord, and furthermore \cite{private} the former decreases while the latter increases upon increasing the total number of quibts $N$ (i.e. enlarging the environment). 

This last result sounds very similar to the findings presented here, and we put forth that both can be understood in terms of the system-environment entanglement. In the system of \cite{Madhok13}, we conjecture that the increase of $N$ results in an increased system-environment entanglement which, despite the increased level of decohering noise, was able to boost the discord in the system. Similarly in our situation, increasing the temperature of the field results in detectors that are more mixed and are thus more entangled with the environment \footnote{In order to use the findings of \cite{KW, Brown13-2} in our argument we must consider an environment that is initially pure. The field is of course not pure when it is thermal. However, we can always consider the field plus its purifying ancilla, and the entanglement between the detectors and this larger environment is quantified simply by the two-detector entropy}. This translates into an increased amount of discord between the detectors.

As a final word on this, we suggest that the evolution of discord in such systems can be understood as a competition between the decohering effect of the environment and the system-environment entanglement that such decoherence also tends to create. As we have seen, oftentimes the latter can win the day. The entanglement within the system, on the other hand, is not bolstered by the system-environment entanglement. Indeed it is will generally be further impaired due to monogamy, and will therefore be far more easily destroyed.

\subsection{Translational invariance}  \label{sectTrans}

In this section we will attempt to give some further mathematical intuition towards the behavior discussed in Sect. \ref{sectResults} by exploiting the translational invariance of the field. In doing so we will uncover a new and interesting perspective on correlation harvesting that we feel is worthy of consideration in its own right. We will also be able to accurately predict for what choices of parameters the thermal amplification phenomenon is strongest, and for which it is weakest.

First let us note that although we use the translational invariance of the field as a convenient means of explanation, this does not imply that translational invariance is necessary for thermal amplification to occur. Indeed it is not, and we have also observed the same effect using a cavity field with mirror boundary conditions instead of periodic.

\begin{figure}[t]
	\centering
    \includegraphics[width=0.45\textwidth]{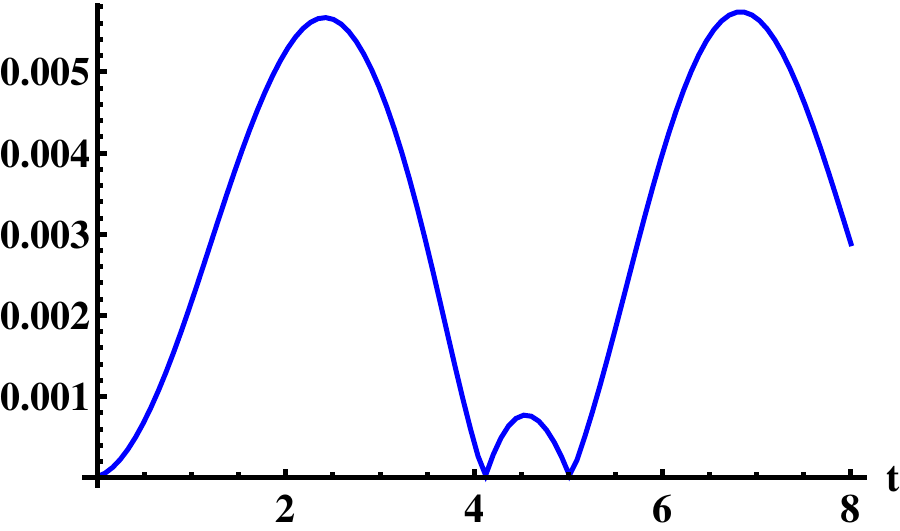}
	\caption{(Color online) The difference $|\nu_1-\nu_2|$ of the symplectic eigenvalues of the detector-detector system as a function of time. The field was initiated in its vacuum state and the detectors were placed a distance $r=4$ away from each other. These are the same parameters that were used to plot the mutual information and discord in Fig. (\ref{vst}-a), and we note that the qualitative behavior in that figure is exactly the same as we see here.}
        \label{nuVt}
\end{figure}

To begin our argument, we will make a simple observation. This is that the mutual information harvested by the detectors appears to be extremely monotonic with the \emph{difference} $|\nu_1-\nu_2|$ of the symplectic eigenvalues of the detector-detector subsystem, given by the covariance matrix $\mat{\sigma}^{(d)}$ in Eq. (\ref{twomode}). For example we can plot $|\nu_1-\nu_2|$ as a function of time of evolution and compare with the extracted mutual information. This is plotted in Fig. (\ref{nuVt}), where the field is initiated in the vacuum state and all parameters are equivalent to those used in Fig. (\ref{vst}). We see that in comparison to the correlation measures plotted in Fig. (\ref{vst}-a), the qualitative behaviors are identical. For large temperatures the the mutual information continues to evolve monotonically with $|\nu_1-\nu_2|$, however this becomes no longer true for discord.

\begin{figure}[t]
	\centering
    \includegraphics[width=0.45\textwidth]{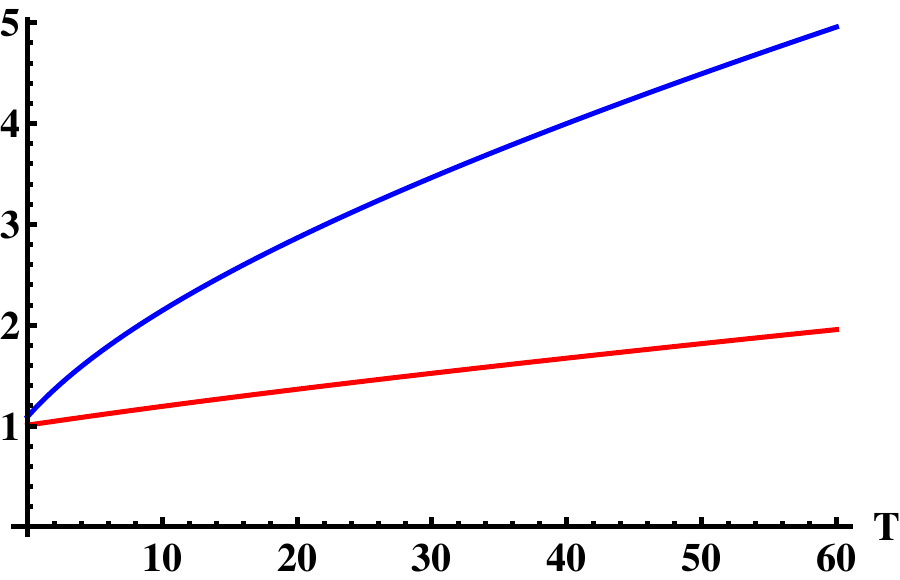}
	\caption{(Color online) The two symplectic eigenvalues $\nu_1$ and $\nu_2$ of the detector-detector system as a function of initial field temperature $T$. The detectors were placed a distance $r=4$ away from each other and left to evolve for a time $t=2$. We see that the difference $|\nu_1-\nu_2|$ grows with temperature.}
        \label{nuVtemp}
\end{figure}

Given this knowledge, we should wonder how $|\nu_1-\nu_2|$ changes with field temperature. Clearly if the field is initiated in a high temperature state we should expect the response of the detectors to be more energetic and for them to become more mixed through their evolution. The symplectic eigenvalues of $\mat{\sigma}^{(d)}$ should therefore increase with larger temperature. This is indeed what occurs, as can be seen in Fig. (\ref{nuVtemp}) where we plot both $\nu_1$ and $\nu_2$ as a function of temperature $T$. However, we also observe that one increases faster than the other, meaning that their difference also grows with $T$. This is therefore consistent with the increase of mutual information $I$ with $T$. Clearly, as can be seen in Fig. (\ref{vsT}), the discord is no longer monotonic for high enough excitation.

The explanation of our results now requires two tasks: first, to explain why the mutual information and (for small temperatures) the discord are monotonic with $|\nu_1-\nu_2|$ and second, to explain why this difference grows with $T$. We will attempt to perform both of these. We will focus on the latter of these to begin, and later give some insight into the former. Being tasked with explaining the behavior seen in Fig. (\ref{nuVtemp}), we will take this as an excuse to introduce an interesting new perspective on the evolution of our system that results from the translational invariance of the field. By this approach we are also very easily, and in a unique manner, able to explain the decay of entanglement seen in Fig. (\ref{entdeg}).

An immediate result of the translational invariance of the field is that both of the detectors will individually feel the exact same response. The detector-detector state is therefore invariant under exchange $1 \leftrightarrow 2$ of the detectors. This implies both that the reduced states of each detector are equivalent $\mat{\sigma}_1 = \mat{\sigma}_2$ (we will therefore refer to both as $\mat{\sigma}_1$ henceforth) and that the off-diagonal correlation matrix is symmetric: $\mat{\gamma}_{12}=\mat{\gamma}_{12}^T$. The covariance matrix of the detectors is therefore of the form
\begin{align}
	\mat{\sigma}^{(d)}=
	\begin{pmatrix}
		\mat{\sigma}_1 & \mat{\gamma}_{1 2} \\
		\mat{\gamma}_{1 2} & \mat{\sigma}_1
	\end{pmatrix}.
\end{align}
This is an example of what is called a symmetric Gaussian state \cite{Adesso07}. From this, it is easily seen that there is a simple symplectic transformation that transforms $\mat{\sigma}^{(d)}$ to a product state, void of any correlation. Of course this is true of any Gaussian state, but in general the correct transformation would depend on the details of the state and, in such a scenario as we have here, would depend on time $t$ and on the chosen parameters. Here, the symmetric form of $\mat{\sigma}^{(d)}$ means that this is not the case, and there is a transformation that will \emph{always} do the job irrespective of time or parameters. We will call this symplectic transformation $\mat{\tilde{S}}$, and it takes the form
\begin{align}   \label{beamsplit}
	\mat{\tilde{S}}=\mat{\tilde{S}}^T=\mat{\tilde{S}}^{-1}=\frac{1}{\sqrt{2}}
	\begin{pmatrix}
		-\mat{I}_2 & \mat{I}_2 \\
		\mat{I}_2 & \mat{I}_2
	\end{pmatrix},
\end{align}
where $\mat{I}_2$ is the $2\times 2$ identity matrix. When applied to the detector-detector state we indeed find that this transformation returns a product state:
\begin{align}  \label{productTrans}
	\mat{\tilde{S}} \, \mat{\sigma}^{(d)}\mat{\tilde{S}}^T=
	\begin{pmatrix}
		\mat{\sigma}_- & \mat{0}_2 \\
		\mat{0}_2 & \mat{\sigma}_+
	\end{pmatrix},
\end{align}
where the single-mode covariance matrices $\mat{\sigma}_{\pm}$ are given by
\begin{align}  \label{beam1}
	\mat{\sigma}_{\pm}=\mat{\sigma}_1 \pm \mat{\gamma}_{12}.
\end{align}
We wish again to emphasize that as long as the detectors start in their ground state then this single transformation will \emph{always} bring our two-detector state to one devoid of correlations, independent of the time of interaction or the distance between them or any other parameters of our system.

Before continuing we wish to point out, as it will be useful later, that the transformation $\mat{\tilde{S}}$ is what's known as a \emph{passive} transformation, due to its orthogonality \cite{Wolf03}. Such transformations preserve the expected energy of the state. In fact, $\mat{\tilde{S}}$ is exactly a $50:50$ beam splitter \cite{Adesso07, Kim02}, an operation that is easily implemented in laboratory settings. The fact that the transformation takes this form will be useful in gaining insights towards the extractable entanglement, and we will discuss this shortly. 

First, we wish to point out that while Eq. (\ref{productTrans}) is easily seen by the detector-exchange symmetry, this result can also be viewed in what is perhaps a more enlightening manner. Recall that the interaction Hamiltonian between the detectors and field, Eq. (\ref{Hint}), is given by
\begin{align}
	\op{H}_\text{int}=\sqrt{2}\lambda \left[\op{q}_{d1}\op{\phi}(x_1)+\op{q}_{d2}\op{\phi}(x_2) \right],
\end{align}
where $\op{q}_{d1}$ and $\op{q}_{d2}$ are the position quadrature operators of the two detectors. This represents a local coupling between detector-1 and $\op{\phi}(x_1)$, and between detector-2 and $\op{\phi}(x_2)$. Of course because the observables $\op{\phi}(x_1)$ and $\op{\phi}(x_2)$ are correlated, as witnessed by the non-vanishing correlation function $C(x_1,x_2) \neq 0$, the two detectors are able to become correlated despite their individual interactions being local. 

On the other hand, we can consider the system under the transformation $\mat{\tilde{S}}$, which when applied to the detector-detector phase space results in
\begin{align}
	\mat{\tilde{S}}
	\begin{pmatrix}
		\hat{q}_{d1} \\
		\hat{p}_{d1} \\
		\hat{q}_{d2} \\
		\hat{p}_{d2} \\
	\end{pmatrix}
	=
	\begin{pmatrix}
		\hat{q}_- \\
		\hat{p}_- \\
		\hat{q}_+ \\
		\hat{p}_+ \\
	\end{pmatrix},
\end{align}
where the new set of quadrature operators are given by
\begin{align}
	\op{q}_\pm=\frac{1}{\sqrt{2}}(\op{q}_{d2} \pm \op{q}_{d1}), \;\;\;\; \op{p}_\pm=\frac{1}{\sqrt{2}}(\op{p}_{d2} \pm \op{p}_{d1}).
\end{align}
We will refer to the corresponding modes as the $(+)$- and $(-)$-modes. It is with respect to these quadratures that the covariance matrices $\mat{\sigma}_\pm$ are defined via Eq. (\ref{covmat}). If we now express the interaction Hamiltonian in this new basis, we see immediately that it is
\begin{align}
	\op{H}_\text{int}=\sqrt{2}\lambda \left[\op{q}_+\op{\phi}_+ +\op{q}_- \op{\phi}_- \right],
\end{align}
where
\begin{align}
	\op{\phi}_\pm = \frac{1}{\sqrt{2}}(\op{\phi}(x_1)\pm \op{\phi}(x_2)).
\end{align}
Again, this represents a pair of local interaction between the $(+)$-mode with $\op{\phi}_+$, and between the $(-)$-mode with $\op{\phi}_-$. However, unlike between $\op{\phi}(x_1)$ and $\op{\phi}(x_2)$, the measurement statistics of the observables $\op{\phi}_+$ and $\op{\phi}_-$ are completely uncorrelated in the vacuum and in thermal states. This can be seen as a direct consequence of translational invariance: for any pair of observables $\op{A}$ and $\op{B}$ that satisfy the $1 \leftrightarrow 2$ exchange symmetry, it is trivially seen that $C(\op{A}_+,\op{B}_-)=0$, where $\op{A}_+=\op{A}(x_1)+\op{A}(x_2)$ and $\op{B}_-=\op{B}(x_1)-\op{B}(x_2)$. This includes of course $C(\op{\phi}_+,\op{\phi}_-)=0$. Thus the $(\pm)$-modes necessarily can never become correlated through the evolution, and are entirely dynamically decoupled, as we have seen. If they are initialized in a correlated state then the mutual information between them can never increase above its initial value (we have confirmed this by direct computation).

What we have found is that the process of correlation extraction (including entanglement) can be fully described by two dynamically decoupled interactions (each between a field and a \emph{single} detector), followed by the beam splitter operation Eq. (\ref{beamsplit}). We are thus able to gain a lot of insight by examining the form of these independent interactions along with the correlating capacity of the beam splitter operation. To proceed, let us first examine the interactions that $\op{q}_+$ and $\op{q}_-$ experience. They are coupled to the operators $\op{\phi}_+$ and $\op{\phi}_-$, respectively, which both take the standard form of a mode-decomposed field:
\begin{align}
	\op{\phi}_\pm=\sum_n \frac{1}{\sqrt{4 \pi |n|}} \left(v_n^{(\pm)} \op{a}_n +v_n^{(\pm) *} \op{a}_n^\dagger \right),
\end{align}
where $\op{a}_n$ and $\op{a}_n^\dagger$ are the same ladder operators as those in Eq. (\ref{field}) (we are not performing a Bogoliubov transformation) and the ``mode functions" are $v_n^{(\pm)}=(\exp (\ii k_n x_2) \pm \exp(\ii k_n x_1))/\sqrt{2}$, or equivalently:
\begin{align} \label{Pmode}
	v_n^{(+)}=\sqrt{2} \cos \left(\frac{k_n}{2} (x_2 - x_1)\right)e^{\ii k_n (x_2+x_1)/2},   \\  \label{Mmode}
	v_n^{(-)}=\sqrt{2}\ii \sin \left(\frac{k_n}{2} (x_2 - x_1)\right)e^{\ii k_n (x_2+x_1)/2}.
\end{align}
Thus, the interactions that the $(\pm)$-modes individually experience can be considered as a standard monopole-monopole coupling to a regular field, except that the effective coupling strengths are frequency- and position-dependent (via the $\cos$ and $\sin$ in the above equations). These interactions will then determine the evolution of the $(\pm)$-modes, and their states as a function of time will be represented by the covariance matrices $\mat{\sigma}_\pm$.

This realization can in fact be used to understand the growth of $|\nu_1-\nu_2|$ seen in Fig. (\ref{nuVtemp}), and therefore indirectly the thermal amplification behavior. Furthermore, by examining $v_n^{(\pm)}$ we are accurately able to predict for which parameter choices thermal amplification will be strongest, and for which it will be weakest. To see this, recall that the symplectic eigenvalues of a Gaussian state are symplectically invariant. It therefore follows that the symplectic eigenvalues $\nu_+$ and $\nu_-$ of the states $\mat{\sigma}_+$ and $\mat{\sigma}_-$ are identified with the original symplectic eigenvalues $\nu_1$ and $\nu_2$ of the detector-detector system. The differences are therefore also equivalent: $|\nu_1-\nu_2|=|\nu_+-\nu_-|$. Fortunately, the qualitative behavior of $|\nu_+-\nu_-|$ is easily predicted by the forms of $v_n^{(\pm)}$ given above. Clearly, if the field state is thermal then we expect the mixedness of the $(\pm)$-modes, and thus both $\nu_+$ and $\nu_-$, to increase with temperature $T$. However, we should also generally expect to find a difference between the two, and this is due to the difference in the magnitudes of $\cos (k_n r/2)$ and $\sin (k_n r/2)$ that appear in the effective coupling strengths. The mode that is coupled more strongly to the field will feel a stronger response, and will be affected more by an increase in temperature, than will the more weakly coupled mode.

For example let us consider the window in time $t$ and distance $r$ that we have examined in the above figures. This is well outside the regime at which a single-mode approximation would be valid; we would need to go to much larger values of $t$. Therefore, roughly speaking, both of the $(\pm)$-modes couple equally to a wide range of modes, not taking account of the $\cos$ and $\sin$ factors. Taking these into account, however, the region of relatively small $r$ that we have examined means that the magnitude of $\sin(k_n r/2)$ is quite small for the many modes of small frequency $k_n$ that are relevant in the evolution. For the same reason, $\cos(k_n r/2)$ for these modes is quite close to unity. We thus find a significant difference in the values of $\nu_+$ and $\nu_-$, with typically $\nu_+ > \nu_-$ for the reasons just stated. If the field is hot, this difference in effective coupling is seen more clearly and thus the thermal amplification of correlations follows. Note that for small $t$ and large $r$ the contribution to the $+$ and $-$ should become roughly equivalent, and thus we expect the strength of thermal amplification to generally fall off with distance. If we allow ourselves to look at larger values of $t$ it turns out that with this framework we are able to accurately predict for which values of $r$ the thermal amplification is strongest, and for which it is weakest. So as not to get too off-track, we present this in the appendix \ref{app}.

Given that we now understand why the difference $|\nu_1-\nu_2|=|\nu_+-\nu_-|$ grows with field temperature, we are left with the task of understanding better why the mutual information $I$, and to a lesser extent the Gaussian discord $D$, are monotonic with this difference. Fortunately we can use the framework just presented to gain some insight on the matter. The behavior of correlations present in $\mat{\sigma}^{(d)}$ of course stem from the correlating properties of the $50:50$ beam splitter $\mat{\tilde{S}}$. For example we can use this fact, along with the well-known entangling properties of beam splitters, to easily explain the entanglement degradation with temperature; this is explained further in the following section. Aside from entanglement, however, there appears to be fairly little information on the beam splitter's ability to generate mutual information and discord in Gaussian states. 

Notice that by inverting Eq. (\ref{beam1}) we obtain
\begin{align}   \label{beam2}
	\mat{\sigma}_1=\mat{\sigma}_2=\frac{1}{2}(\mat{\sigma}_++\mat{\sigma}_-), \;\;\;\; \mat{\gamma}_{12}=\frac{1}{2}(\mat{\sigma}_+-\mat{\sigma}_-).
\end{align}
Clearly if the states of the $(\pm)$-modes are the same, $\mat{\sigma}_+=\mat{\sigma}_-$, then the difference in symplectic eigenvalues is zero and the detector-detector state is completely uncorrelated since $\mat{\gamma}_{12}=\mat{0}$. That is, $\mat{\tilde{S}}$ takes an uncorrelated pair of identical Gaussian states and outputs exactly the same thing. On the other hand, the larger the difference between $\mat{\sigma}_+$ and $\mat{\sigma}_-$, the larger the correlation matrix $\mat{\gamma}_{12}$ will be. Of course an increase in this difference does not necessarily correspond to an increase in $|\nu_+-\nu_-|$. However, in our particular scenario the forms that we obtain for $\mat{\sigma}_+$ and $\mat{\sigma}_-$ tend to be approximately thermal, $\mat{\sigma}_\pm \approx \text{diag}(\nu_\pm, \nu_\pm)$, at least for relatively small temperatures of the field. For exactly thermal states, one has exactly $\mat{\gamma}_{12}=\text{diag}(\nu_+-\nu_-,\nu_+-\nu_-)/2$, and thus in this case the magnitude of the correlation matrix does directly correspond with the magnitude $|\nu_+-\nu_-|=|\nu_1-\nu_2|$. This provides the qualitative explanation that we were searching for.

Of course the mutual information and discord do not depend purely on $\mat{\gamma}_{12}$, and they will both decrease as the overall mixedness of the system is increased. In fact, if we continue with the above approximation of the states $\mat{\sigma}_{\pm}$ both being exactly thermal, one finds that both $I$ and $D$ are monotonically increasing with $|\nu_+-\nu_-|$, but also monotonically decreasing with $\nu_++\nu_-$. The extraction of correlations therefore represents a competition between these two quantities. For the Gaussian discord it is seen that the sum $\nu_+ + \nu_-$ plays a stronger role than in the mutual information. The discord is thus more sensitive to noise (as we have observed) but for small enough field temperatures $T$ (yet still very large) this sensitivity is not enough to overcome the increase in $|\nu_+-\nu_-|$ achieved by increasing $T$.

We can display this in a more quantitative manner by plotting $I$ as a function of the symplectic eigenvalues. Given Eq. (\ref{beam2}), we see that the mutual information, Eq. (\ref{mutinf}) takes the form
\begin{align}
	I=2f\left(\frac{1}{2}\sqrt{\det(\mat{\sigma}_++\mat{\sigma}_-)} \right)-f(\nu_+)-f(\nu_-),
\end{align}
where $f$ is the function given by Eq. (\ref{ffunction}). In the case that $\mat{\sigma}_\pm$ are exactly thermal, and taking into account the identification of $\{\nu_+,\nu_-\}$ with $\{\nu_1,\nu_2\}$, this simplifies to
\begin{align}  \label{mutinf2}
	I=2f\left(\frac{1}{2}(\nu_1+\nu_2) \right)-f(\nu_1)-f(\nu_2).
\end{align}
\begin{figure}[t]
	\centering
        \includegraphics[width=0.38\textwidth]{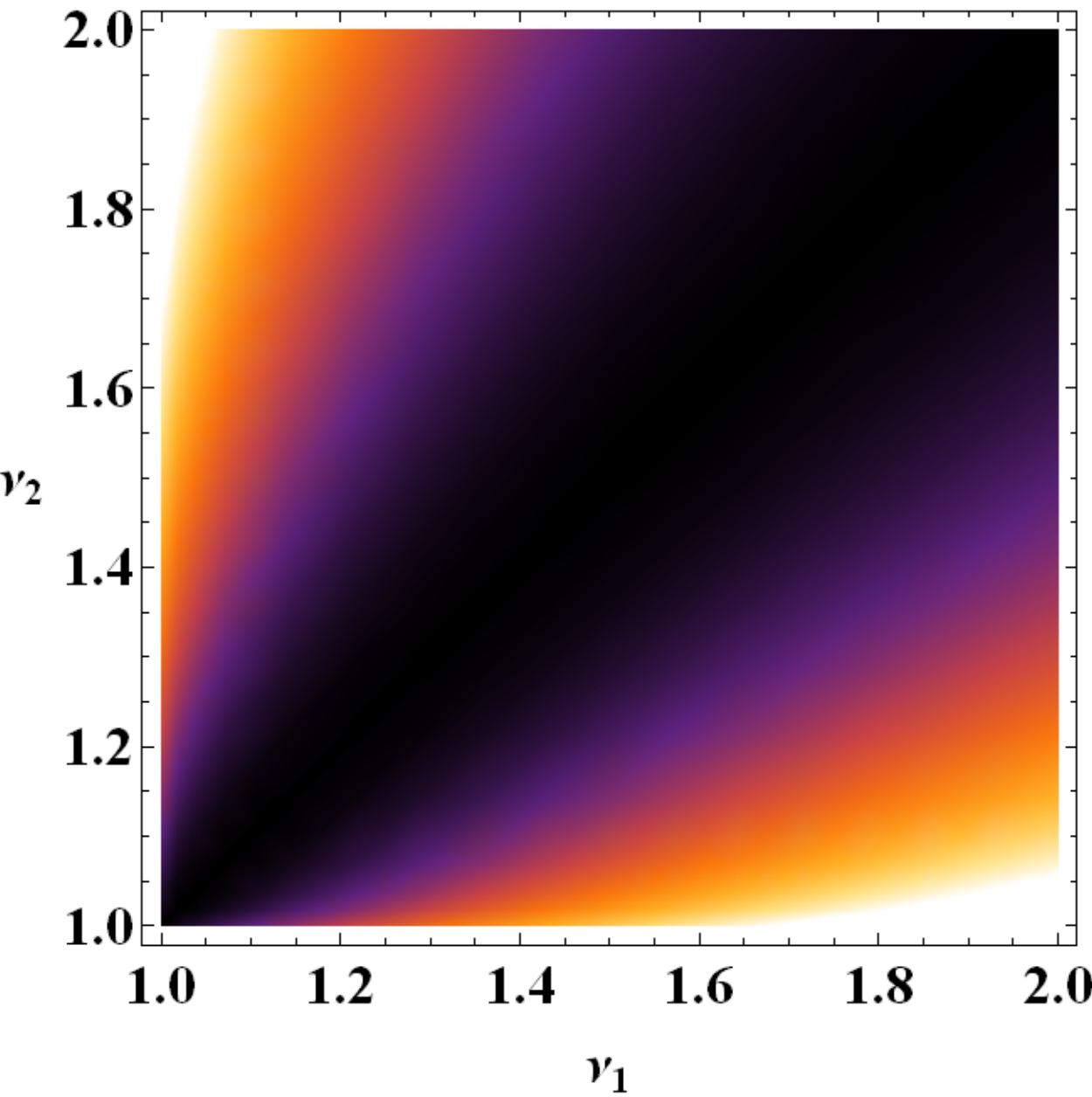}
	\includegraphics[width=0.09 \textwidth]{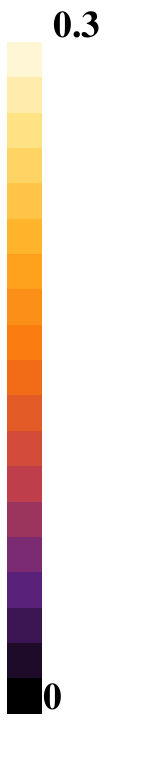}
	\caption{(Color online) The mutual information, Eq. (\ref{mutinf2}), as a function of $\nu_1$ and $\nu_2$. We see clearly the increase with $|\nu_1-\nu_2|$.}
        \label{density2}
\end{figure}
We plot this in Fig. (\ref{density2}). We see clearly in this figure the increase in $I$ as the difference $|\nu_+-\nu_-|$ increases. The corresponding plot for Gaussian discord looks nearly identical to this, except that the functional decrease with $\nu_1+\nu_2$ is more dominant over the increase with $|\nu_1-\nu_2|$.

\subsection{Entanglement degradation}

It is worth noting that with the framework presented above we can immediately explain the degradation of extracted entanglement with temperature, as seen in Fig. (\ref{entdeg}), via the known entangling properties of passive operations, which the $50:50$ beam splitter represented by $\mat{\tilde{S}}$ is an example of. This transformation happens to be its own inverse, and thus applying it to the product state $\mat{\sigma}_- \oplus \, \mat{\sigma}_+$ returns back the detector-detector state, $\mat{\sigma}^{(d)}$. It is known that for passive operations to create entanglement the original state must be ``nonclassical", \cite{Kim02}. What this actually means generically and quantitatively is that an entangling passive operation exists iff the two smallest eigenvalues $\lambda_1$ and $\lambda_2$ (\emph{not} the symplectic eigenvalues) of the original covariance matrix satisfy $\lambda_1 \lambda_2 <1$ \cite{Wolf03}, and furthermore this product is used to provide a maximal amount of entanglement that can be achieved. Thus, if we can understand from the evolution of the $(\pm)$-modes how the eigenvalues of $\mat{\sigma}_- \oplus \, \mat{\sigma}_+$ change with field temperature, then we are able to garner information on the entanglement that can be present in the state $\mat{\sigma}^{(d)}$.

Due to $\mat{\sigma}_- \oplus \, \mat{\sigma}_+$ being a product state, the eigenvalues of this matrix will of course just be the combination of the eigenvalues of the two individual $(\pm)$-modes. These are generically of the form $\nu_- e^{r_-}$, $\nu_- e^{-r_-}$, $\nu_+ e^{r_+}$, and $\nu_+ e^{-r_+}$. Here $\nu_-$ and $\nu_+$ are just the symplectic eigenvalues presented above, and the values $r_-$ and $r_+$ are the single-mode squeezing parameters for each of the two modes \cite{Adesso07}. Note that with large enough squeezing it becomes very easy to entangle via passive operations because the two smallest eigenvalues will be $\nu_- e^{-r_-}$ and $\nu_+ e^{-r_+}$, and they will become very small as $r_+$ and $r_-$ become large. In our scenario the evolution does provide some amount of squeezing (it must, in order to get any entanglement at all). However, the symplectic eigenvalues $\nu_\pm$ are what generically increase when the field temperature is increased (being directly related to the mixedness of the modes) and thus as field temperature is increased the values of $\nu_\pm$ will quickly overtake the squeezing such that the two smallest eigenvalues no longer satisfy $\lambda_1 \lambda_2<1$. Once this inequality is broken, it is no longer possible to obtain any entanglement in $\mat{\sigma}^{(d)}$.

It is interesting to note, however, that for the case of a product state (as we have) in which one of the modes is thermal (as we approximately have) then the \emph{maximally} entangling passive operation is in fact a $50:50$ beam splitter \cite{Wolf03}. In a way then, the field is actually doing the best it possibly can to entangle the detectors. Unfortunately, despite this great effort, thermal fluctuations quickly win the day.

\section{Conclusions}   \label{conclusions}

Using the oscillator-detector formalism presented in \cite{Brown13} we have non-perturbatively and exactly solved for the local harvesting of classical and quantum correlations (entanglement, Gaussian discord, and mutual information) from a cavity scalar field. We have furthermore explored the effect that thermal fluctuations in the field will have on the harvesting of these correlations. As expected, the harvested entanglement rapidly decays with the field temperature. Surprisingly, however, both the mutual information as well as the Gaussian discord (a measure of purely quantum correlations) can be greatly \emph{increased} by increasing the field temperature. Indeed, an improvement of multiple orders of magnitude is possible in this regard.

Although initially surprising, we go on to discuss that this result can be physically understood in several different ways. We have included an explanation of thermal amplification in terms of the field correlation function, as well as speculated on its possible relation to system-environment entanglement. Our primary explanation of the phenomenon relies on the translational invariance inherent in the periodic field. By this we are able to explain both the results of thermal degradation of entanglement as well as the thermal amplification of mutual information and discord purely in terms of the correlating capability of the $50:50$ beam splitting operation. We can furthermore use this to accurately predict for what system parameters thermal amplification will be weakest, and for which it will be strongest. We feel that, independent of its explanatory power that has been demonstrated here, the new perspective that we present on correlation extraction is interesting and worthy of consideration in its own right. This is because in general the evolution of a single detector with a field is much more easily understood intuitively than is the collective behavior of two detectors. The ability to decompose our two-detector scenario into two dynamically decoupled, single-detector systems represents a tool for gaining intuition into harvesting and similar phenomena.

We note that the phenomenon of thermal amplification does not require translational invariance in order to occur.  Indeed, we have also observed thermal amplification using a cavity field with mirror boundary conditions, as would be the case in an actual optical cavity. In fact, we  propose that thermal amplification may be a rather general phenomenon that can occur in a variety of quantum systems (for example other physical realizations of a collection harmonic oscillators, such as in condensed matter physics). The particular scenario that we have considered here may be just one instance of a much broader phenomenon.

The thermal amplification of discord appears to imply the possibility of non-locally generating what is an appreciable amount of quantum correlations. This is exciting both from the experimental perspective as well as, perhaps, practical discord-based quantum computation. For example, it is known that discord allows the local activation of entanglement with an ancilla system \cite{Piani11,Streltsov11,Adesso13}. Interestingly then, even though thermal fluctuations are directly detrimental to entanglement, they may nevertheless be used as an indirect tool for its generation. This may moreover prove useful due to the fact that many of the significant technological hurdles facing quantum computation stem from the necessity to keep one's system very cold. Our results suggest that in a specific scenario one may need not worry about thermal noise in their system, and indeed may even welcome it. Assuming that we are able to fully understand thermal amplification more generally (in all its possible physical realizations) and how to properly utilize the resources that it produces, this may very well give way to a type of ``noise-assisted quantum computation".

\section{Acknowledgments}
Much thanks must be given to Eduardo Mart\'{i}n-Mart\'{i}nez and Robert Mann for their helpful discussions and for reviewing this work prior to publication, as well as to Aida Ahmadzadegan for her excellent proofreading. We also thank Gerardo Adesso for pointing out an important reference that we had overlooked.
This work has been supported in part by the National Sciences and Engineering
Research Council of Canada.

\appendix

\section{Making predictions} \label{app}

\begin{figure}[t]
	\centering
        \includegraphics[width=0.38\textwidth]{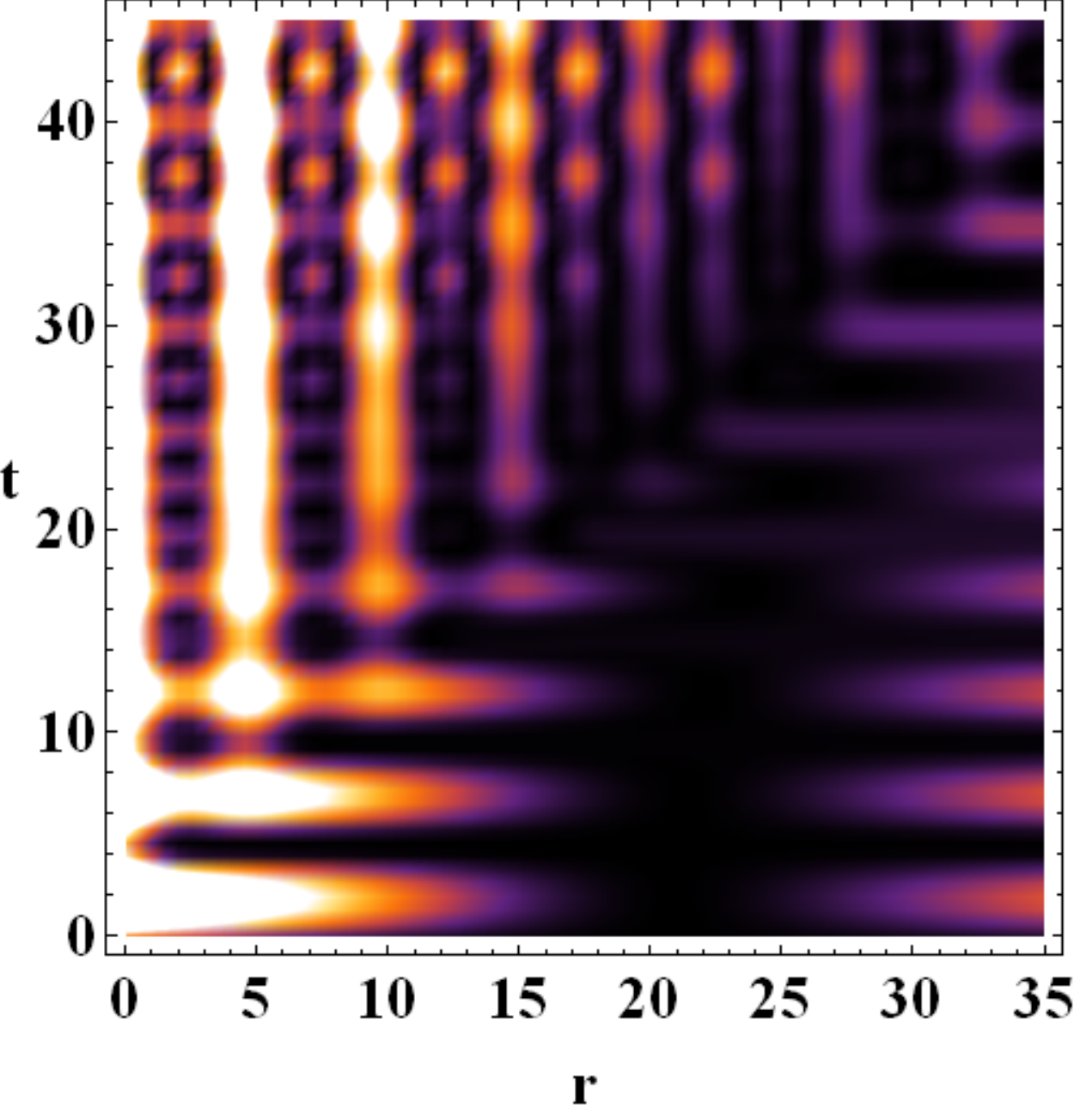}
	\includegraphics[width=0.09 \textwidth]{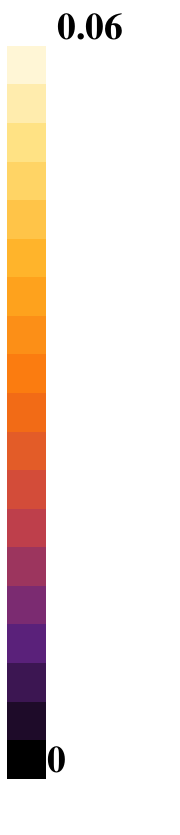}
	\caption{(Color online) The mutual information between detectors as a function of $t$ and $r$, where the field was initiated in a thermal state of temperature $T=2$.}
        \label{bigdensity}
\end{figure}

Here we wish to point out that the material discussed in Sects. \ref{sectCorr} and \ref{sectTrans} can actually be used to make accurate predictions about how strongly we expect to observe the thermal amplifications for varying parameters. Displayed in Fig. (\ref{bigdensity}) is the extracted mutual information  as a function of $t$ and $r$ in the case that the field temperature was initiated to be $T=2$; all other parameters are as they were in Sect. \ref{sectResults}. Note that a similar plot of the Gaussian discord looks qualitatively almost identical, but with a somewhat reduced magnitude.

Notice that there is a clear transition at the light cone $t=r$. In the spacelike region, $t<r$, we see that there is a minimum at roughly $r \approx 21$. This value of $r$ coincides with a local minimum (a zero, in fact) of the magnitude of the correlation function $C(r)$ discussed in Sect. \ref{sectCorr}. Inside the light cone, however, when $t>r$, we see that this no longer plays a strong role. This is simply because for $t>r$ the detectors have come into causal contact, and thus the generation of correlations can also follow from the direct exchange of quanta rather than the harvesting of correlations from the field. Interestingly, the behavior of $I$ and $D$ in the $t<r$ region is qualitatively very similar to the behavior found in the vacuum, but for $t>r$ the behavior is very different (in the case of the vacuum, there is no significant change in behavior across the light cone aside from a visible amplification near $t=r$ due the exchange of real quanta).

In the region of causal contact, $t>r$, the behavior of $I$ and $D$ is very different from the spacelike region, and it is here that we observe the intensity of thermal amplification following directly from the couplings of the $(\pm)$-modes to the field, Eqs. (\ref{Pmode}, \ref{Mmode}). In this region we are starting to see resonant effects. This means that we expect to see the strongest contribution to the evolution of the detectors coming from the field modes near the resonance frequency $|k_n|=\Omega=40 \pi/L$, with $L=100$. For example we notice the bands of strong amplification at distances of $r=0,5,10\dots$. From what we have learned in the main text, these should correspond to large differences in the values of the symplectic eigenvalues $\nu_+$ and $\nu_-$. Recalling that the $(\pm)$-modes are coupled to different field modes with strengths that go as $\cos(k_n r/2)$ and $\sin(k_n r/2)$, we note that $r=0,5,10\dots$ are exactly the values that satisfy $\Omega r/2=m \pi$ for integer values of $m$, meaning that the coupling of the $(\pm)$-modes to the resonant frequency are $|\cos(\Omega r/2)|=1$ and $|\sin(\Omega r/2)|=0$. This maximum difference then leads to a large difference in the symplectic eigenvalues, and thus to a local maximum in the thermal amplification.

We also notice in Fig. (\ref{bigdensity}) that there are local maxima at values of $r=2.5, 7.5,12.5 \dots$; these of course are the values that satisfy $\Omega r/2=m \pi/2$ and thus give $|\cos(\Omega r/2)|=0$ and $|\sin(\Omega r/2)|=1$. This again translates into a maximum in the difference $|\nu_+-\nu_-|$ and thus in thermal amplification. However, why are these maxima much weaker than the others discussed above? This follows from the same reasoning that was discussed in the main text: given the timescale we are observing, we are still far from the single-mode approximation being accurate. There are thus significant contributions coming from the field modes $k_n$ of small frequency, and for relatively small $r$ this translates into $\sin(k_n r/2)$ being very small for most of these modes. Thus, for small $t$ and small $r$ we generically have $\nu_+ > \nu_-$. In this regime the maxima of $\nu_-$ are not able to overcome the minima of $\nu_+$, and thus we do not see local maxima in the value of $|\nu_+-\nu_-|$. However, as $t$ increases we should expect these maxima to emerge and become more prominent, and this is indeed what we observe in Fig. (\ref{bigdensity}). In the large $t$ limit the maxima at $r=2.5,7.5,12.5 \dots$ should be equivalent to those at $r=0,5,10\dots$ as the single-mode approximation becomes accurate. Furthermore, we see that for large values of $r$ the two sets of maxima become comparable, which is also consistent with our discussion.


\begin{thebibliography}{10}

\bibitem{Horodecki09}
R.~Horodecki, P.~Horodecki, M.~Horodecki, and K.~Horodecki,
\newblock Rev. Mod. Phys {\bf 81}, 865 (2009).

\bibitem{Ollivier01}
H.~Ollivier and W.~H. Zurek,
\newblock Phys. Rev. Lett {\bf 88}, 017901 (2001).

\bibitem{Henderson01}
L.~Henderson and V.~Vedral,
\newblock J. Phys. A: Math. Theor {\bf 34}, 6899 (2001).

\bibitem{Modi12}
K.~Modi, A.~Brodutch, H.~Cable, T.~Paterek, and V.~Vedral,
\newblock Rev. Mod. Phys {\bf 84}, 1655 (2012).

\bibitem{Girolami13}
D.~Girolami, T.~Tufarelli, and G.~Adesso,
\newblock Phys. Rev. Lett {\bf 110}, 240402 (2013).

\bibitem{Streltsov13}
A.~Streltsov and W.~H. Zurek,
\newblock Phys. Rev. Lett {\bf 111}, 040401 (2013).

\bibitem{Brown13-2}
E.~G. Brown, E.~J. Webster, E. Mart\'{i}n-Mart\'{i}nez, and A.~Kempf,
\newblock Annals of Physics. {\bf 337}, 153 (2013).

\bibitem{Madhok13}
V.~Madhok, V.~Gupta, A.~M. Hamel, and S.~Ghose,
\newblock arXiv:1307.1405 [quant-ph]  (2013).

\bibitem{Knill98}
E.~Knill and R.~Laflamme,
\newblock Phys. Rev. Lett {\bf 81}, 5672 (1998).

\bibitem{Dakic10}
B.~Dakic, V.~Vedral, and C.~Brukner,
\newblock Phys. Rev. Lett {\bf 105}, 190502 (2010).

\bibitem{Chuan12}
T.~K. Chuan {\em et~al.},
\newblock Phys. Rev. Lett {\bf 109}, 070501 (2012).

\bibitem{Madhok13-2}
V.~Madhok and A.~Datta,
\newblock Int. J. Mod. Phys. B {\bf 27}, 1245041 (2013).

\bibitem{Almedia13}
M.~de~Almeida {\em et~al.},
\newblock arXiv:1301.7110 [quant-ph]  (2013).

\bibitem{Gu13}
M.~Gu {\em et~al.},
\newblock Nat. Phys {\bf 8}, 671 (2013).

\bibitem{Piani11}
M.~Piani {\em et~al.},
\newblock Phys. Rev. Lett {\bf 106}, 220403 (2011).

\bibitem{Streltsov11}
A.~Streltsov, H.~Kampermann, and D.~Bruss,
\newblock Phys. Rev. Lett {\bf 106}, 160401 (2011).

\bibitem{Adesso13}
G.~Adesso, V.~D'Ambrosio, E.~Nagali, M.~Piani, and F.~Sciarrino,
\newblock arXiv:1308.1680 [quant-ph]  (2013).

\bibitem{Adesso10}
G.~Adesso and A.~Datta,
\newblock Phys. Rev. Lett {\bf 105}, 030501 (2010).

\bibitem{Giorda10}
P.~Giorda and M.~G.~A. Paris,
\newblock Phys. Rev. Lett {\bf 105}, 020503 (2010).

\bibitem{Reznik05}
B.~Reznik, A.~Retzker, and J.~Silman,
\newblock Phys. Rev. A {\bf 71}, 042104 (2005).

\bibitem{Braun05}
D.~Braun,
\newblock Phys. Rev. A {\bf 72}, 062324 (2005).

\bibitem{Leon09}
J.~Leon and C.~Sabin,
\newblock Phys. Rev. A {\bf 79}, 012304 (2009).

\bibitem{Steeg09}
G. Ver Steeg and N.~C. Menicucci,
\newblock Phys. Rev. D {\bf 79}, 044027 (2009).

\bibitem{Summers87}
S.~J. Summers and R.~Werner,
\newblock Commun. Math. Phys {\bf 110}, 247 (1987).

\bibitem{Srednicki93}
M.~Srednicki,
\newblock Phys. Rev. Lett {\bf 71}, 666 (1993).

\bibitem{Edu2013}
E. Mart\'{i}n-Mart\'{i}nez, E.~G. Brown, W.~Donnelly, and A.~Kempf,
\newblock arXiv:1309.1090 [quant-ph]  (2013).

\bibitem{Borrelli12}
M.~Borrelli, C.~Sabin, G.~Adesso, F.~Plastina, and S.~Maniscalco,
\newblock New Journal of Physics {\bf 14}, 103010 (2012).

\bibitem{Brown13}
E.~G. Brown, E. Mart\'{i}n-Mart\'{i}nez, N.~C. Menicucci, and R.~B. Mann,
\newblock Phys. Rev. D {\bf 87}, 084062 (2013).

\bibitem{Dragan11}
A.~Dragan and I.~Fuentes,
\newblock arXiv:1105.1192 [quant-ph]  (2011).

\bibitem{Bruschi13}
 D. E.~Bruschi, A.~R.~Lee and I.~Fuentes,
\newblock J. Phys. A: Math. Theor {\bf 46}, 165303 (2013).

\bibitem{Sabin11}
C.~Sabin, M.~del Rey, J.~J. Garcia-Ripoll, and J.~Leon,
\newblock Phys. Rev. Lett {\bf 107}, 150402 (2011).

\bibitem{Sabin12}
C.~Sabin, B.~Peropadre, M.~del Rey, and E. Mart\'{i}n-Mart\'{i}nez,
\newblock Phys. Rev. Lett {\bf 109}, 033602 (2012).

\bibitem{Strel}
A. ~Streltsov, H. ~Kampermann, and D. ~Bruss
\newblock Phys. Rev. Lett {\bf 107}, 170502 (2011)

\bibitem{Anders08}
J.~Anders and A.~Winter,
\newblock Quantum Information and Computation {\bf 8}, 0245 (2008).

\bibitem{Ferraro08}
A.~Ferraro, D.~Cavalcanti, A.~Garcia-Saez, and A.~Acin,
\newblock Phys. Rev. Lett. {\bf 100}, 080502 (2008).

\bibitem{Werlang09}
T.~Werlang, S.~Souza, F.~F. Fanchini, and C.~J. Villas Boas,
\newblock Phys. Rev. A {\bf 80}, 024103 (2009).

\bibitem{Wang2010}
B.~Wang, Z.-Y. Xu, Z.-Q. Chen, and M.~Feng,
\newblock Phys. Rev. A {\bf 81}, 014101 (2010).

\bibitem{Datta2009}
A.~Datta,
\newblock Phys. Rev. A {\bf 80}, 052304 (2009).

\bibitem{Brown2012}
E.~G. Brown, K.~Cormier, E. Mart\'{i}n-Mart\'{i}nez, and R.~B. Mann,
\newblock Phys. Rev. A {\bf 86}, 032108 (2012).

\bibitem{Doukas2013}
J.~Doukas, E.~G. Brown, A.~Dragan, and R.~B. Mann,
\newblock Phys. Rev. A {\bf 87}, 012306 (2013).

\bibitem{Paz}
J. N. Freitas and J. P. Paz,
\newblock Phys. Rev. A {\bf 85}, 032118 (2012).

\bibitem{Adesso07}
G.~Adesso and F.~Illuminati,
\newblock J. Phys. A: Math. Theor {\bf 40}, 7821 (2007).

\bibitem{BirrelDavies84}
N.~D. Birrel and P.~C.~W. Davies,
\newblock {\em Quantum Fields in Curved Space} (Cambridge University Press,
  1984).

\bibitem{Schumaker86}
B.~L. Schumaker,
\newblock Physics Reports {\bf 135}, 317 (1986).

\bibitem{DeWitt80}
B.~DeWitt,
\newblock {\em General Relativity; an Einstein Centenary Survey} (Cambridge
  University Press, 1980).

\bibitem{Plenio05}
M.~B. Plenio,
\newblock Phys. Rev. Lett {\bf 95}, 090503 (2005).

\bibitem{Giorda12}
P.~Giorda, M.~Allegra, and M.~G.~A. Paris,
\newblock Phys. Rev. A {\bf 86}, 052328 (2012).

\bibitem{Weldon00}
H.~A. Weldon,
\newblock Phys. Rev. D. {\bf 62}, 056010 (2000).

\bibitem{Wolf03}
M.~M. Wolf, J.~Eisert, and M.~B. Plenio,
\newblock Phys. Rev. Lett {\bf 90}, 047904 (2003).

\bibitem{Kim02}
M.~S. Kim, W.~Son, V.~Buzek, and P.~L. Knight,
\newblock Phys. Rev. A {\bf 65}, 032323 (2002).

\bibitem{KW}
F. F. Fanchini, M. F. Cornelio, M. C. de Oliveira, and A. O. Caldeira,
\newblock Phys. Rev. A {\bf 84}, 012313 (2011).

\bibitem{private}
V.~Gupta,
\newblock Private communication .

\end{thebibliography}
\end{document}